\documentclass[superscriptaddress,
twocolumn,
]{revtex4-1}
\usepackage[dvipdfmx
]{graphicx}
\usepackage{amsmath}
\usepackage{bm}
\begin{document}

\title{%
Photoinduced dynamics of excitonic order and Rabi oscillation in the 
two-orbital Hubbard model}
\date{\today}

\author{Yasuhiro Tanaka}
\email{tanaka.y@aoni.waseda.jp}
\affiliation{Department of Physics, Tokyo Institute of Technology, Meguro, Tokyo 152-8550, Japan}
\affiliation{Department of Applied Physics, Waseda University, Shinjuku, Tokyo 169-8555, Japan}
\thanks{Present address}
\author{Kenji Yonemitsu}
\affiliation{Department of Physics, Chuo University, Bunkyo, Tokyo 112-8551, Japan}

\begin{abstract}
We investigate the condition for the photoinduced enhancement 
of an excitonic order in a two-orbital Hubbard model, 
which has been theoretically proposed in our previous work 
[Phys. Rev. B 97, 115105 (2018)], and analyze it from 
the viewpoint of the Rabi oscillation. Within the mean-field approximation, 
we simulate real-time dynamics of an 
excitonic insulator with a direct gap, where the pair condensation in the 
initial state is of 
BEC nature and the photoexcitation is introduced by electric dipole transitions. 
We first discuss that in the atomic limit our
 model is reduced to a two-level 
system that undergoes the Rabi oscillation, so that for single cycle pulses 
physical quantities after the photoirradiation are essentially determined by 
the ratio of the Rabi frequency to the pump-light frequency. Then, it is shown
 that this picture holds even in the case of nonzero transfer 
integrals where each one-particle state exhibits the Rabi oscillation leading 
to the enhancement of the excitonic order. We demonstrate that effects of 
electron-phonon interactions do not alter the results qualitatively. We also examine 
many-body dynamics by the exact diagonalization method on small clusters, which 
strongly suggests that our mechanism for the enhancement of the exctionic order
survives even when quantum fluctuations are taken into account. 

\end{abstract}

\maketitle

\section{Introduction}
Photoirradiation to correlated electron systems has opened a novel 
playground to manipulate various electronic phases. 
In particular, recent experimental studies have reported that 
electronic orders are transiently reinforced or even created by 
laser light \cite{Onda_PRL08,Fausti_Sci11,Ishikawa_NatComm14,Hu_NatMat14,
Kaiser_PRB14,Stoj_SCI14,Mitrano_Nat16,Singer_PRL16,Mor_PRL17,Kawakami_PRB17}, 
which indicates a clear distinction from typical photoinduced phase 
transitions in which they are usually suppressed. 
These phenomena have been observed, for instance, in 
materials which exhibit 
charge ordering \cite{Onda_PRL08,Ishikawa_NatComm14,Kawakami_PRB17}, 
charge density wave \cite{Stoj_SCI14,Singer_PRL16}, 
superconductivity \cite{Fausti_Sci11,Hu_NatMat14,Kaiser_PRB14,
Mitrano_Nat16}, and excitonic condensation \cite{Mor_PRL17}. 
Simultaneously, theoretical efforts to understand their mechanisms 
as well as to pursue a way of controlling electronic phases have 
been made recently, where roles of electron-electron (e-e) and/or 
electron-phonon (e-ph) interactions on laser-induced dynamics have been 
intensively studied \cite{Lu_PRL12,Tsuji_PRB12,Hashimoto_JPSJ13,Hashimoto_JPSJ14,
Yanagiya_JPSJ15,Nakagawa_PRL15,Yonemitsu_JPSJ17,Ido_SCAD17,Murakami_PRL17,
Tanaka_PRB18,Oya_PRB18,Tanabe_PRB18}. 
For excitonic insulators (EIs), a transient gap enhancement by 
photoexcitation has been observed in a candidate material 
Ta$_2$NiSe$_5$ \cite{Mor_PRL17}. The EI is a state in which 
electrons in the conduction band and holes in the valence band 
form bound pairs called excitons by the Coulomb interaction, and 
they become a condensate. Theories of EIs have been 
developed in semimetals and 
semiconductors \cite{Mott_PM61,Knox_SSP63,Jerome_PR67,Halperin_RMP68,Kunes_JPCM15}. 
Ta$_2$NiSe$_5$ is a layered semiconductor with a direct gap 
above $T_C=326$K where a second-order transition accompanied by 
a structural distortion occurs \cite{Wakisaka_PRL09,Salvo_JLCM86}. 
Although the identification of an EI is a difficult task, 
recent experimental \cite{Lu_NatComm17,Li_PRB18} and 
theoretical \cite{Seki_PRB14,Sugimoto_PRB16,Matsuura_JPSJ16,Sugimoto_PRL18} 
studies have offered evidences that an EI is realized in the 
low temperature phase. With regards to its photoinduced phenomena,  
e-ph coupled systems have been investigated by 
mean-field theories \cite{Murakami_PRL17,Tanabe_PRB18} and the origin of 
the gap enhancement has been discussed.  

In purely electronic systems without phonon degrees of freedom, 
we have studied \cite{Tanaka_PRB18} photoinduced 
dynamics of a direct-gap EI using a two-orbital Hubbard model 
in which excitonic 
condensation in thermal equilibrium shows BCS-BEC 
crossover depending on the value of the interorbital Coulomb 
interaction $U^{\prime}$. By incorporating the effects of 
photoexcitation through electric dipole transitions, we have 
shown that the enhancement of the excitonic gap occurs when the 
initial state 
is an EI in the BEC regime or a nearby band insulating state, 
and the pump-light frequency is close to the excitonic gap. 
There is an optimal value of the amplitude of the light field for 
inducing the gap enhancement, although its physical origin has 
not been clarified yet. Our study has also shown that the time 
evolutions of the phases of excitonic pairs in momentum space 
are crucially important for understanding the 
photoinduced behavior of the excitonic gap \cite{Tanaka_PRB18}: 
They evolve basically in phase when the gap is enhanced by the 
laser irradiation, 
whereas they strongly depend on momentum when the initial EI is 
in the BCS regime for which the gap is suppressed. 

In this paper, we elucidate the physical origin of the 
gap enhancement through laser-induced dipole transitions in 
a two-orbital Hubbard model mainly by using the time-dependent 
Hartree-Fock (HF) approximation. 
For this purpose, we consider the atomic limit in which our 
system is equivalent to a two-level system that exhibits the Rabi 
oscillation, where the dynamics of physical quantities are 
understood from changes in the occupation probability of the two levels. 
Even when we introduce nonzero transfer integrals, its photo-response 
is qualitatively unaltered as far as 
the initial state is near the boundary between the EI and the 
band insulator (BI) phases. There the EI belongs to the BEC regime and 
the excitonic pairs are formed locally. 
In momentum space, the photoinduced gap enhancement is interpreted as a 
consequence of a cooperative Rabi oscillation of the one-particle states. 
We confirm that the e-ph coupling considered in the previous theories 
\cite{Murakami_PRL17,Tanabe_PRB18} has little effects on our mechanism 
for the gap enhancement. 
Moreover, we examine effects of quantum fluctuations on the dynamics by using 
the exact diagonalization (ED) method, which corroborates the results
 obtained by the HF approximation. This 
paper is organized as follows. In Sect. II, the two-orbital Hubbard model and the 
calculation method for photoinduced dynamics are introduced. The model in 
the atomic limit is also described.  In Sect. III, 
the results without the phonon degrees of freedom are presented and we discuss 
the photoinduced dynamics in terms of the Rabi oscillation. The effects of the e-ph 
coupling are elucidated in Sect. IV, whereas those of quantum fluctuations are 
discussed in Sect. V where we give the results with the ED method. 
The discussion and summary are devoted to Sect. VI. 

\section{Model and Method}
\subsection{Two-orbital Hubbard model}
We consider a two-orbital Hubbard model in one dimension, which is defined as
\begin{eqnarray}
\hat{H}&=&t_{c}\sum_{i\sigma}(c^{\dagger}_{i\sigma}c_{i+1\sigma}+h.c.)+\mu_C\sum_{i\sigma}n^{c}_{i\sigma} \nonumber \\
&+&t_{f}\sum_{i\sigma}(f^{\dagger}_{i\sigma}f_{i+1\sigma}+h.c.)\nonumber \\
&+&U\sum_{i}n^{c}_{i\uparrow}n^{c}_{i\downarrow}+U\sum_{i}n^{f}_{i\uparrow}n^{f}_{i\downarrow}+U^{\prime}\sum_{i}n^{c}_{i}n^{f}_{i},
\label{eq:ham}
\end{eqnarray}
where $\alpha^{\dagger}_{i\sigma}$ and $\alpha_{i\sigma}$ ($\alpha=c$, $f$) are creation and 
annihilation operators for an electron with spin $\sigma$ ($=\uparrow, \downarrow)$ at 
the $i$th site on the $\alpha$ orbital, respectively. 
The number operators are defined by $n^{\alpha}_{i\sigma}=\alpha^{\dagger}_{i\sigma}\alpha_{i\sigma}$ 
and $n^{\alpha}_i=n^{\alpha}_{i\uparrow}+n^{\alpha}_{i\downarrow}$. 
The intraorbital (interorbital) Coulomb interaction is denoted by $U$ ($U^{\prime}$). 
For the transfer integral $t_{\alpha}$, we set $t_f=1$ and $t_c=-1$ as in the previous 
study \cite{Tanaka_PRB18}. 
The parameter $\mu_C(>0)$ controls the overlap between the $c$ and $f$ bands. 
When $\mu_C>4$, the system with $U=U^{\prime}=0$ has a band structure of a 
direct-gap semiconductor, whereas it becomes a semimetal for $\mu_C<4$. 
The electron density per site is fixed at $n=2$. 

Photoexcitation is introduced by electric dipole-allowed 
transitions \cite{Golez_PRB16,Murakami_PRL17} that are described by the 
time ($\tau$)-dependent term 
\begin{equation}
\hat{H}_D(\tau)=F(\tau)\sum_{i\sigma}(c^{\dagger}_{i\sigma}f_{i\sigma}+h.c.), 
\label{eq:ham_D}
\end{equation}
which is added to Eq. (\ref{eq:ham}). We define $F(\tau)$ as
\begin{equation}
F(\tau)=F_0\sin(\omega \tau)e^{-(\tau-\tau_p)^2/\tau_w^2}, 
\label{eq:fg_tau}
\end{equation}
where $\omega$ is the light frequency and we set $\tau_p=\tau_w=\pi/\omega$. 
Although we mainly use the  
gaussian envelope for $F(\tau)$, we also consider a rectangular 
envelope with which $F(\tau)$ is defined as
\begin{equation}
F(\tau)=F_0\sin(\omega \tau)\theta (\tau)\theta (T_{\rm irr}-\tau), 
\label{eq:fs_tau}
\end{equation}
where $\theta(\tau)$ and $T_{\rm irr}$ are the Heaviside step function and the pulse width, 
respectively. This form of $F(\tau)$ enables us to interpret our results directly from 
the viewpoint of the 
Rabi oscillation. We note that the pulse shape does not qualitatively affect our 
results. 
Unless otherwise noted, we use single cycle pulses ($T_{\rm irr}=2\pi /\omega$). 

We apply the HF approximation to Eq. (\ref{eq:ham}) where 
the excitonic order parameter and the electron density on the 
$\alpha$ orbital per site are defined as
$\Delta_0=\langle c^{\dagger}_{i\sigma}f_{i\sigma}\rangle$ and 
$n_{\alpha}=2\langle n^{\alpha}_{i\sigma}\rangle$, respectively. 
We have assumed that $\langle c^{\dagger}_{i\sigma}f_{i\sigma}\rangle$ and 
$\langle n^{\alpha}_{i\sigma}\rangle$ are independent of $i$ and 
$\sigma$ \cite{Tanaka_PRB18,comment1}. 
The total Hamiltonian in momentum representation is given as 
\begin{equation}
\hat{H}^{\rm HF}_{\rm tot}(\tau)=\sum_{k\sigma}\hat{H}_{k\sigma}(\tau)=\sum_{k\sigma}\Psi^{\dagger}_{k\sigma}h_{k}(\tau)\Psi_{k\sigma},
\label{eq:ham_t}
\end{equation}
where $\Psi^{\dagger}_{k\sigma}=(c^{\dagger}_{k\sigma}, f^{\dagger}_{k\sigma})$ 
and $h_{k}(\tau)$ is defined by
\begin{eqnarray}
h_{k}(\tau)=
\left(\begin{array}{cc} \tilde{\epsilon}^c_{k} & -U^{\prime}\Delta^{\ast}_0+F(\tau)\\ 
-U^{\prime}\Delta_0+F(\tau) & \tilde{\epsilon}^f_{k}\\ \end{array} \right).
\label{eq:ham_mat}
\end{eqnarray}
In Eq. (\ref{eq:ham_mat}), 
$\tilde{\epsilon}^c_{k}=\epsilon^c_{k}+\frac{U}{2}n_c+U^{\prime}n_f$ and 
$\tilde{\epsilon}^f_{k}=\epsilon^f_{k}+\frac{U}{2}n_f+U^{\prime}n_c$ where 
$\epsilon^c_{k}=2t_c\cos k+\mu_C$ and $\epsilon^f_{k}=2t_f\cos k$ are the 
noninteracting energy dispersions for the $c$ and $f$ bands, respectively. 
In the ground state, $n_c$ ($=2-n_f$) and $\Delta_0$ are determined 
self-consistently. 

Photoinduced dynamics are obtained by numerically solving the time-dependent 
Schr$\ddot{\rm o}$dinger equation \cite{Terai_TPS93,Kuwabara_JPSJ95,Tanaka_JPSJ10,comment2}
\begin{equation}
|\psi_{k\sigma}(\tau+d\tau)\rangle = T\exp \Bigl[ -i\int^{\tau+d\tau}_{\tau}d\tau^{\prime}\hat{H}_{k\sigma}(\tau^{\prime})\Bigr]
|\psi_{k\sigma}(\tau)\rangle, 
\label{eq:schr_eq}
\end{equation}
where $|\psi_{k\sigma}(\tau)\rangle$ denotes a 
one-particle state with wave vector $k$ and spin $\sigma$ at time $\tau$, and 
$T$ is the time-ordering operator. 
We use the time slice $d\tau=0.01$ with $t_f$ as the unit of energy 
(and $1/t_f$ as that of time). 
For a physical quantity $X(\tau)$, its time average is denoted by $\overline{X}$ 
that is calculated as 
\begin{equation}
\overline{X}=\frac{1}{\tau_f-\tau_i}\int^{\tau_f}_{\tau_i}X(\tau)d\tau. 
\label{eq:time_av}
\end{equation}
If $X(\tau)$ is conserved after the photoexcitation, its value is written as
$\widetilde{X}$. 

\subsection{Atomic limit}
In the atomic limit ($t_c=t_f=0$), our system is reduced to a two-level 
system described by the Hamiltonian 
\begin{equation}
\hat{H}^{\rm AL}(\tau)=\sum_{\sigma}\Psi^{\dagger}_{\sigma}h^{\rm AL}(\tau)\Psi_{\sigma}, 
\end{equation}
where $\Psi^{\dagger}_{\sigma}=(c^{\dagger}_{\sigma},f^{\dagger}_{\sigma})$ and $h^{\rm AL}(\tau)$ 
is defined as
\begin{eqnarray}
h^{\rm AL}(\tau)=
\left(\begin{array}{cc} \epsilon^c & -U^{\prime}\Delta^{\ast}_0+F(\tau)\\ 
-U^{\prime}\Delta_0+F(\tau) & \epsilon^f\\ \end{array} \right), 
\label{eq:ham_mat_al}
\end{eqnarray}
with $\epsilon^c=\mu_C+\frac{U}{2}n_c+U^{\prime}n_f$ and 
$\epsilon^f=\frac{U}{2}n_f+U^{\prime}n_c$. In the ground state [$F(\tau)=0$], 
the self-consistent equations for $n_c$ and $\Delta_0$ are written as
\begin{equation}
n_c=1-\frac{\epsilon^c-\epsilon^f}{\sqrt{(\epsilon^c-\epsilon^f)^2+4{U^{\prime}}^2|\Delta_0|^2}}, 
\end{equation}
and
\begin{equation}
\Delta_0=\frac{1}{2}\Biggl[ 1-\frac{(\epsilon^c-\epsilon^f)^2}{(\epsilon^c-\epsilon^f)^2+4{U^{\prime}}^2|\Delta_0|^2}\Biggr]^{1/2}, 
\end{equation}
respectively, which leads to 
\begin{equation}
\Delta_0=\frac{1}{2}\sqrt{n_c(2-n_c)}.
\label{eq:delta-nc}
\end{equation}
With this relation, the expectation value for the energy $E^{\rm AL}$ can be 
written as 
\begin{equation}
E^{\rm AL}=\frac{1}{2}(U-U^{\prime})n_c^2+(\mu_C-U+U^{\prime})n_c+U. 
\label{eq:gs_ener_al}
\end{equation}
We take $U$ ($1/U$) as the unit of energy (time) in the atomic limit. 

\section{Results Without Phonons}

In this section, we show the results obtained by the HF approximation for 
the two-orbital Hubbard model without e-ph couplings. We first consider  
the case of the atomic limit and then discuss the case of nonzero transfer 
integrals ($t_f=-t_c=1$).

\subsection{Atomic limit}

\subsubsection{Ground state}
\begin{figure}
\includegraphics[height=7.5cm]{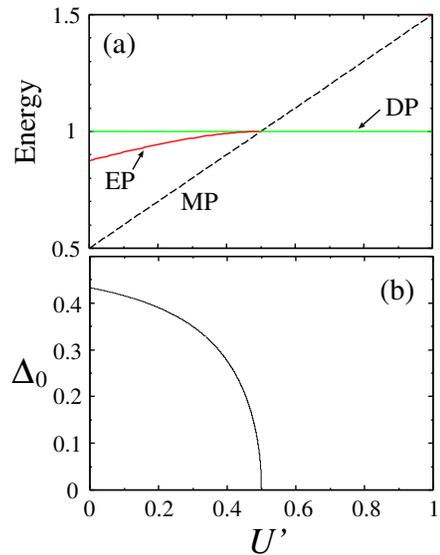}
\caption{(a) Ground-state energies for EP and DP, and (b) $\Delta_0$ as 
functions of $U^{\prime}$ with $U=1$ and $\mu_C=0.5$. 
In (a), the energy of the magnetic phase (MP) is also shown by 
the dashed line.}
\label{fig:fig1}
\end{figure}

Before the laser irradiation, we consider two phases in the ground state: 
an excitonic phase (EP) with $\Delta_0\neq 0$ and a decoupled phase (DP) 
with $\Delta_0=0$ and $n_c=0$. 
They correspond to EI and BI phases, 
respectively, when $t_c$ and $t_f$ are nonzero \cite{Kaneko_PRB12,Tanaka_PRB18}. 
We use $U=1$ and $\mu_C=0.5$, and vary $U^{\prime}$ ($\geq 0$) as a parameter.  
Since $\partial{E^{\rm AL}}/\partial{n_c}=0$ gives 
$n_c=1-\mu_C/(U-U^{\prime})$, we have a mean-field solution 
with $\Delta_0\neq 0$ for $0\leq U^{\prime}<U^{\prime}_{\rm cr}=U-\mu_C$. 
In Fig. \ref{fig:fig1}(a), we show the ground-state energies for the EP and 
the DP, the latter of which is independent of $U^{\prime}$. 
With increasing $U^{\prime}$, a transition from the EP to the DP occurs 
at $U^{\prime}=U^{\prime}_{\rm cr}$. 
In fact, a magnetic phase (MP) with $\langle n^c_{\sigma}\rangle =1$ and 
$\langle n^f_{-\sigma}\rangle =1$ has the ground-state energy $U^{\prime}+\mu_C$, 
which  
becomes the lowest energy state for $0\leq U^{\prime}<U^{\prime}_{\rm cr}$. 
However, we consider 
only nonmagnetic initial states for the photoexcitation. 
The reason is that for nonzero $t_c$ and $t_f$, the photoinduced gap 
enhancement reported previously occurs in the vicinity of the 
boundary between the EI and BI phases \cite{Tanaka_PRB18} 
($U^{\prime}\sim U^{\prime}_{\rm cr}$) where 
 magnetic ordered states do not 
appear as the ground state \cite{Zocher_PRB11}. 
In this paper, we focus on the dynamics near $U^{\prime}=U^{\prime}_{\rm cr}$. 

In Fig. \ref{fig:fig1}(b), the $U^{\prime}$ dependence of $\Delta_0$ is 
shown, indicating that $\Delta_0$ is nonzero at $U^{\prime}=0$ and 
exhibits a steep decrease toward $U^{\prime}=U^{\prime}_{\rm cr}$ at which it 
vanishes. 
For $U^{\prime}\sim U^{\prime}_{\rm cr}$, this result is 
similar to that obtained with nonzero $t_c$ and $t_f$ 
(Fig. \ref{fig:fig10} in Sect. III B), whereas they are qualitatively 
different for $U^{\prime}\sim 0$. 
The similarity near $U^{\prime}=U^{\prime}_{\rm cr}$ comes from the local 
character of excitonic pairs: the EI is in the BEC regime of the BCS-BEC 
crossover \cite{Kaneko_PRB12,Phan_PRB10,Seki_PRB11}. 
On the other hand, for $U^{\prime}\sim 0$, the EI is in the BCS regime, 
which cannot be described by the atomic limit. 

\subsubsection{Photoinduced dynamics}
\begin{figure}
\includegraphics[height=7.5cm]{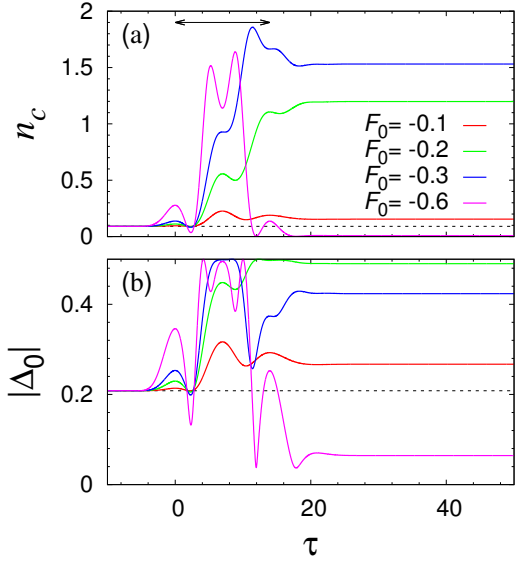}
\caption{Time evolutions of (a) $n_c$ and (b) $|\Delta_0|$ for different 
values of $F_0$ with $U=1$, $\mu_C=0.5$, $U^{\prime}=0.45$, and $\omega=0.45$. 
The double-headed arrow indicates the range $0<\tau<2\tau_w=2\pi/\omega$ of 
application of an electric field. The horizontal 
dashed line in each panel indicates the corresponding equilibrium value.}
\label{fig:fig2}
\end{figure}

In Fig. \ref{fig:fig2}, we show the time evolutions of $n_c$ and 
$|\Delta_0|$ for different values of $F_0$ 
with $U^{\prime}=0.45$ ($<U^{\prime}_{\rm cr}$). For the time evolutions of the 
real and imaginary parts of $\Delta_0$, see Appendix A.
We use $F(\tau)$ with the gaussian envelope [Eq. (\ref{eq:fg_tau})] and choose $F_0<0$, 
although the sign of $F_0$ does not affect the results 
qualitatively \cite{Tanaka_PRB18}. The pump-light frequency 
is tuned to the difference between the two 
eigenvalues of Eq. (\ref{eq:ham_mat_al}) with $F(\tau)=0$. 
Since $n_c$ and $|\Delta_0|$ are conserved after the photoexcitation, 
they are denoted by $\widetilde{n_c}$ and $\widetilde{|\Delta_0|}$, respectively. 
As we increase $|F_0|$, $\widetilde{n_c}$ and $\widetilde{|\Delta_0|}$ 
become larger than those in the ground state. 
When $|F_0|$ is increased further ($|F_0|=0.6$), they are smaller than 
those at $\tau=0$. This behavior is qualitatively the same as that 
obtained in the previous study for the two-dimensional model near 
the EI-BI phase boundary \cite{Tanaka_PRB18}. 

\begin{figure}
\includegraphics[height=12.5cm]{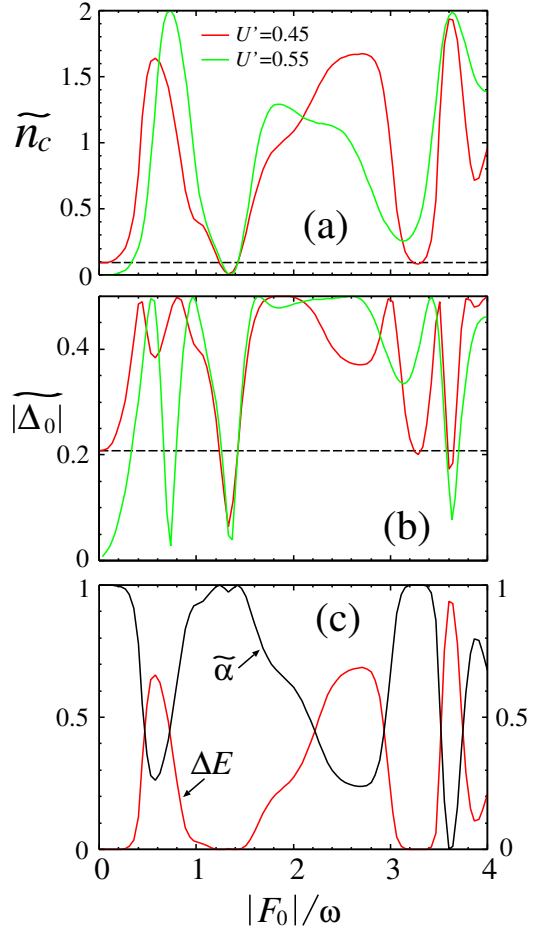}
\caption{(a) $\widetilde{n_c}$, (b) $\widetilde{|\Delta_0|}$, and 
(c) $\Delta E$ and $\widetilde{\alpha}$ as functions of $|F_0|/\omega$ 
for $U=1$ and $\mu_C=0.5$. We use $F(\tau)$ with the gaussian envelope 
defined in Eq. (\ref{eq:fg_tau}). In (a) and (b), the results with 
$U^{\prime}=0.45$ and $0.55$ are shown, where the pump-light frequencies 
are $\omega=0.45$ and $\omega=0.6$, respectively. The dashed horizontal lines 
indicate the corresponding equilibrium values for $U^{\prime}=0.45$. 
For $U^{\prime}=0.55$, $n_c$ and $|\Delta_0|$ are zero in equilibrium. 
In (c), only the results with $U^{\prime}=0.45$ are shown.} 
\label{fig:fig3}
\end{figure}

In Figs. \ref{fig:fig3}(a) and \ref{fig:fig3}(b), we show 
$\widetilde{n_{c}}$ and 
$\widetilde{|\Delta_0|}$ as functions of $|F_0|/\omega$ for $U^{\prime}=0.45$ 
and $0.55$, where the ground states are in the EP and the DP, respectively. 
At time $\tau$, the wave function of the two-level system is 
written as 
\begin{equation}
|\psi(\tau)\rangle =u(\tau)c^{\dagger}|0\rangle
+v(\tau)f^{\dagger}|0\rangle, 
\label{eq:one-particle}
\end{equation}
with $|u(\tau)|^2+|v(\tau)|^2=1$ where we have omitted 
the spin index for brevity. By using the relations 
$n_c=2|u(\tau)|^2$ and $\Delta_0=u^{\ast}(\tau)v(\tau)$, we have 
\begin{equation}
|\Delta_0|=\frac{1}{2}\sqrt{n_c(2-n_c)}, 
\label{eq:delta-nc}
\end{equation}
which holds at any $\tau$ indicating that $\widetilde{|\Delta_0|}$ has its maximum 
value of $0.5$ when $\widetilde{n_c}=1$. 
In order to examine changes in the occupation probability of the two 
levels, we compute 
the overlap between the wave function in the ground 
state and that after the photoexcitation. 
The overlap $\alpha$ is given by
\begin{eqnarray}
\alpha&=&|\langle \psi(\tau=0)|\psi(\tau)\rangle|^2\nonumber \\
&=& 1-\frac{1}{2}(n_c+n^G_c)+\frac{1}{2}n_cn^G_c+{\Delta_0^G}^{\ast}\Delta_0
+\Delta_0^G{\Delta_0}^{\ast}, 
\label{eq:alpha}
\end{eqnarray}
where $n^G_c$ and $\Delta_0^G$ are $n_c$ and $\Delta_0$ in the ground state, 
respectively. 
In evaluating $\alpha$, we adjust the phase of $\Delta_0$ in $|\psi(\tau=0)\rangle$
to coincide with that of $\Delta_0(\tau \neq 0)$. In this case, we have 
\begin{equation}
\alpha = 1-\frac{1}{2}(n_c+n^G_c)+\frac{1}{2}n_cn^G_c+2|\Delta_0^G||\Delta_0|, 
\label{eq:alpha2}
\end{equation}
which is conserved after the photoexcitation. 
In Fig. \ref{fig:fig3}(c), we show $\widetilde{\alpha}$ and the 
increment in the total energy $\Delta E$ per site for $U^{\prime}=0.45$. 
For the quantities $\widetilde{n_c}$, $\Delta E$, and $\widetilde{\alpha}$, 
an oscillatory behavior with respect to $|F_0|/\omega$ is evident, although 
the period of oscillation is not constant. The 
behavior of $\widetilde{|\Delta_0|}$ appears to be more complex than 
that of $\widetilde{n_c}$ because of the relation Eq. (\ref{eq:delta-nc}). 
The oscillation in $\widetilde{\alpha}$ and $\Delta E$ 
indicates a manifestation of the Rabi oscillation \cite{Rabi_PR37,Allen_BooK} 
in the present two-level system, which we will discuss in detail 
below. 

\begin{figure}
\includegraphics[height=12.5cm]{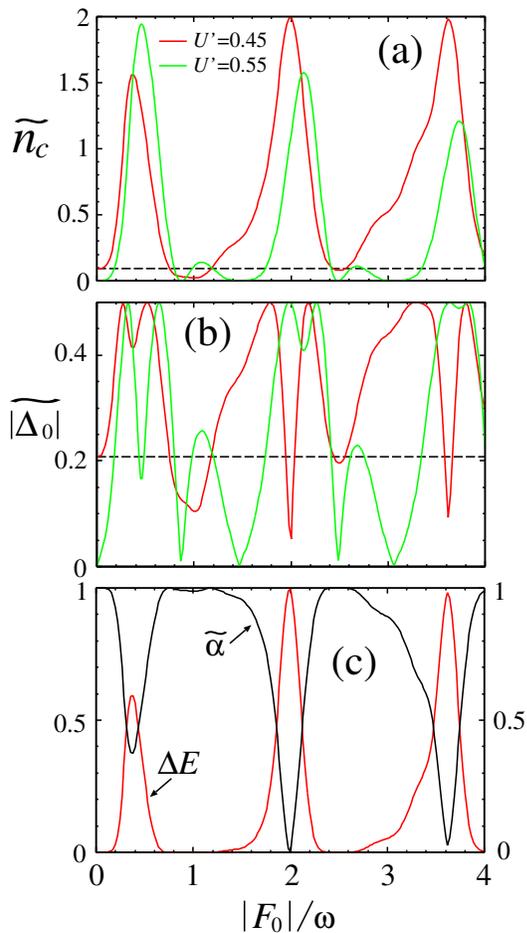}
\caption{Same plots as Fig. \ref{fig:fig3} except that we use $F(\tau)$ with the 
rectangular envelope defined by Eq. (\ref{eq:fs_tau}).} 
\label{fig:fig3-sin}
\end{figure}

\subsubsection{Rabi oscillation and enhancement of excitonic order}
Here we consider $F(\tau)$ with the rectangular envelope 
defined by Eq. (\ref{eq:fs_tau}). In Fig. \ref{fig:fig3-sin}, 
we show  $\widetilde{n_c}$, $\widetilde{|\Delta_0|}$, $\Delta E$, and
 $\widetilde{\alpha}$ as functions of $|F_0|/\omega$ where the parameters 
are the same as those in Fig. \ref{fig:fig3}. The oscillatory 
behavior of these quantities is more prominent than that in Fig. \ref{fig:fig3} 
where the gaussian envelope is employed for $F(\tau)$.
In particular, the period of the oscillation is almost constant. 
\begin{figure}
\includegraphics[height=5.0cm]{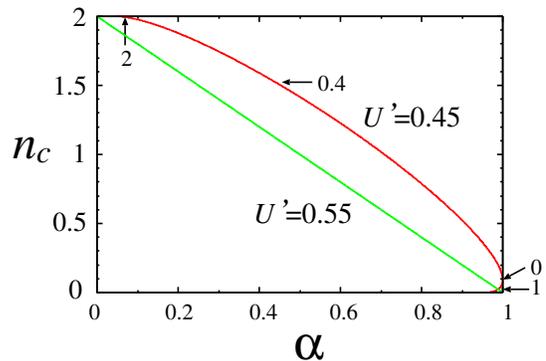}
\caption{Relation between $\alpha$ and $n_c$ for $U^{\prime}=0.45$ and $0.55$. 
For $U^{\prime}=0.45$, the position $(\alpha, n_c)$ at $|F_0|/\omega=0$ 
and $(\widetilde{\alpha}, \widetilde{n_c})$ with $|F_0|/\omega=0.4$, $1$, 
and $2$ are depicted by the arrows.}
\label{fig:fig4}
\end{figure}
By using Eqs. (\ref{eq:delta-nc}) and (\ref{eq:alpha2}), we obtain the 
$\alpha$ dependence of $n_c$ shown in Fig. \ref{fig:fig4}. 
Along its curve, the position $(\widetilde{\alpha}, \widetilde{n_c})$ 
moves depending on the value of $|F_0|/\omega$. 
For $U^{\prime}=0.45$, we have $(\alpha, n_c)=(1, 0.091)$ in the 
ground state. 
With increasing $|F_0|/\omega$, the position 
($\widetilde{\alpha}$, $\widetilde{n_c}$) 
first moves to the upper-left direction in Fig. \ref{fig:fig4} until 
$|F_0|/\omega \sim 0.4$ where 
$\widetilde{n_c}$ exhibits 
the first peak as shown in Fig. \ref{fig:fig3-sin}(a). 
Reflecting the periodic behavior of $\widetilde{\alpha}$,  
the point ($\widetilde{\alpha}$, $\widetilde{n_c}$) goes back to 
the initial position at $|F_0|/\omega \sim 0.8$. 
The value of $\widetilde{n_c}$ becomes smaller than 
$n^G_c$ for $0.8\lesssim |F_0|/\omega\lesssim 1.2$ 
[Fig. \ref{fig:fig3-sin}(a)] where $\widetilde{\alpha}$ is slightly smaller than 1
[Fig. \ref{fig:fig3-sin}(c)]. 
Then, ($\widetilde{\alpha}$, $\widetilde{n_c}$) moves to the upper-left 
direction again until $|F_0|/\omega=2.0$ at which $\widetilde{n_c}$ 
shows the second peak. 
For $U^{\prime}=0.55$, Eq. (\ref{eq:alpha2}) gives 
$\alpha=1-\frac{1}{2}n_c$ since $n^G_c=\Delta^G_0=0$. The behavior of 
($\widetilde{\alpha}$, $\widetilde{n_c}$) depending on $|F_0|/\omega$ 
is similar to that for $U^{\prime}=0.45$. 
These results show that the oscillatory behavior of physical quantities 
originates from that of $\widetilde{\alpha}$. 
In order to interpret our results as the Rabi oscillation more 
quantitatively, we consider the case of 
continuous-wave (CW) lasers in the following. 

\begin{figure}
\includegraphics[height=5.0cm]{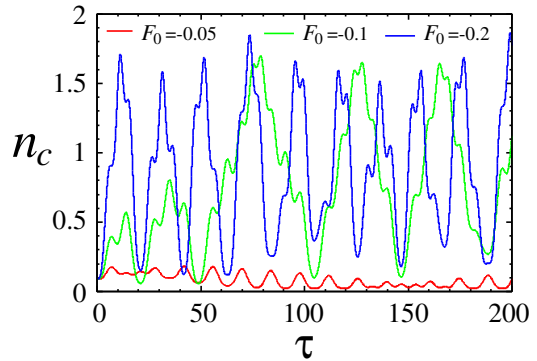}
\caption{Time evolution of $n_c$ under CW excitations for 
different values of $F_0$. We use $U=1$, $\mu_C=0.5$, $U^{\prime}=0.45$, and 
$\omega=0.45$.}
\label{fig:fig5}
\end{figure}

In Fig. \ref{fig:fig5}, we show the time profile of $n_c$ under 
CW excitations for $U^{\prime}=0.45$ and $\omega=0.45$. 
When $|F_0|$ is small ($F_0=-0.05$), $n_c$ shows a small 
oscillation around the value of $n_c^G$. As we increase $|F_0|$, 
a large-amplitude oscillation appears, 
the period of which gets shorter for larger $|F_0|$. 
\begin{figure}[h]
\includegraphics[height=5.0cm]{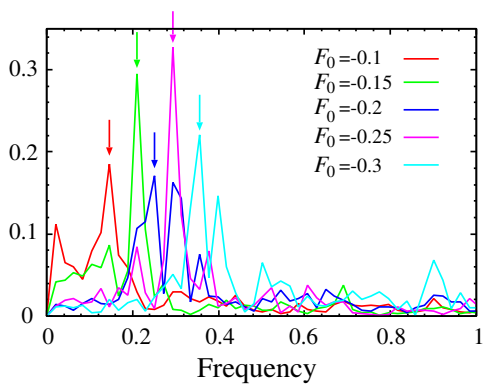}
\caption{Fourier transform of $n_c$. 
In each spectrum, the position of its largest peak is indicated by the 
arrow. The parameters are the same as those in Fig. \ref{fig:fig5}.}
\label{fig:fig6}
\end{figure}
In Fig. \ref{fig:fig6}, we show the Fourier transform of $n_c$ for large 
$|F_0|$ ($F_0\leq -0.1$) (see Appendix A for the details of the dynamics 
for small $|F_0|$). 
There is a sharp peak in each spectrum and its position denoted by 
$\Omega$ is nearly proportional to $|F_0|$ as shown in Fig. \ref{fig:fig7}. 
In a two-level system driven by a CW laser, the rotating 
wave approximation (RWA) gives the Rabi frequency $\Omega_R$ as 
\begin{figure}[h]
\includegraphics[height=5.0cm]{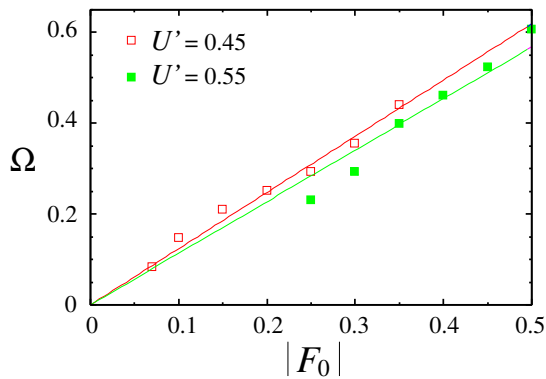}
\caption{Peak frequency $\Omega$ in Fourier transform of $n_c$ as a 
function of $|F_0|$ for $U^{\prime}=0.45$ and $0.55$. 
The solid lines are fitting results.}
\label{fig:fig7}
\end{figure}
\begin{equation}
\Omega_R = \sqrt{(\omega-E_G)^2+{F_0}^2}, 
\end{equation}
where $E_G$ is the difference between the two energy levels. 
At the resonance ($\omega=E_G$), we have $\Omega_R=|F_0|$. 
In fact, the Hamiltonian in Eq. (\ref{eq:ham_mat_al}) contains 
$n_c$ and $\Delta_0$ that are $\tau$ dependent, which is different from the 
conventional Rabi oscillation \cite{Rabi_PR37,Allen_BooK}. 
The effects of the $\tau$-dependence of $n_c$ and $\Delta_0$ in the 
Hamiltonian on the dynamics are discussed in 
Appendix B. Considering this difference, here we will replace $\Omega_R$ at the 
resonance [$\omega=E_G(\tau=0)$] by $\Omega^{\prime}_R=p|F_0|$ with a 
coefficient $p$ \cite{Ono_PRB16}. This leads us to 
$u(\tau)\propto \sin(\frac{\Omega^{\prime}_R}{2}\tau+\phi)$ 
\begin{figure}[h]
\includegraphics[height=5.4cm]{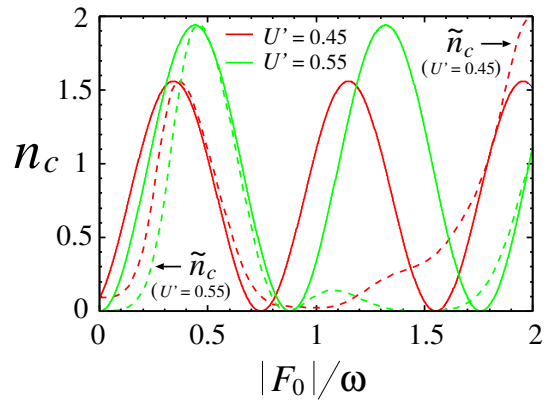}
\caption{$|F_0|/\omega$ dependence of $n_c$ obtained by Eq. (\ref{eq:nc-rabi}) at 
$\tau=2\pi/\omega$. We show $\widetilde{n_c}$ in Fig. \ref{fig:fig3-sin}(a) by 
the dashed lines for comparison. We use $U^{\prime}=0.45$ and $0.55$.}
\label{fig:fig8}
\end{figure}
so that $n_c$ is written as
\begin{equation}
n_c=A\sin^2(\frac{\Omega^{\prime}_R}{2}\tau+\phi), 
\label{eq:nc-rabi}
\end{equation}
where $A$ and $\phi$ are constants. 
We fit a linear function $\Omega=p|F_0|$ to the results of $\Omega$ in 
Fig. \ref{fig:fig7}. 
The fitting works well with $p$ slightly larger than $1$ for 
both $U^{\prime}=0.45$ ($p=1.24$) and $0.55$ ($p=1.14$). 
In Eq. (\ref{eq:nc-rabi}), we have $\phi\sim 0$ ($\phi = 0$) 
when $U^{\prime}$ is slightly smaller (larger) than $U^{\prime}_{\rm cr}$ 
because of $n^G_c\sim 0$ ($n^G_c=0$), indicating that 
for single cycle pulses the values of $n_c$ and $|\Delta_0|$ 
after the photoexcitation are governed by $\Omega^{\prime}_R/\omega$. 
In particular, $n_c$ becomes maximum at $\Omega^{\prime}_R/\omega\sim 1/2$. 
This relation gives $|F_0|/\omega=0.40$ 
for $U^{\prime}=0.45$ and $|F_0|/\omega=0.44$ for $U^{\prime}=0.55$, 
which are consistent with the results shown 
in Fig. \ref{fig:fig3-sin}(a). 
In Fig. \ref{fig:fig8}, we show the $|F_0|/\omega$ dependence of $n_c$ 
calculated by Eq. (\ref{eq:nc-rabi}) at $\tau=2\pi/\omega$, 
and compare the result with $\widetilde{n_c}$ 
shown in Fig. \ref{fig:fig3-sin}(a). The quantities 
$A$ and $\phi$ are determined from the height of the first 
peak in $\widetilde{n_c}$ and the value of $n^G_c$. 
For $|F_0|/\omega \lesssim 0.8$, 
the results obtained by Eq. (\ref{eq:nc-rabi}) reproduce 
those for $\widetilde{n_c}$ fairly well, although they deviate 
from each other for larger 
$|F_0|/\omega$, which is due to the limitation of 
the RWA \cite{Nishioka_JPSJ14}. 

\subsection{One-dimensional model}

Next, we show results for the one-dimensional 
model with $t_f=1$ and $t_c=-1$ in Eq. (\ref{eq:ham}), for 
which the initial EI and BI have a direct gap \cite{Tanaka_PRB18}. 
\begin{figure}[h]
\includegraphics[height=4.5cm]{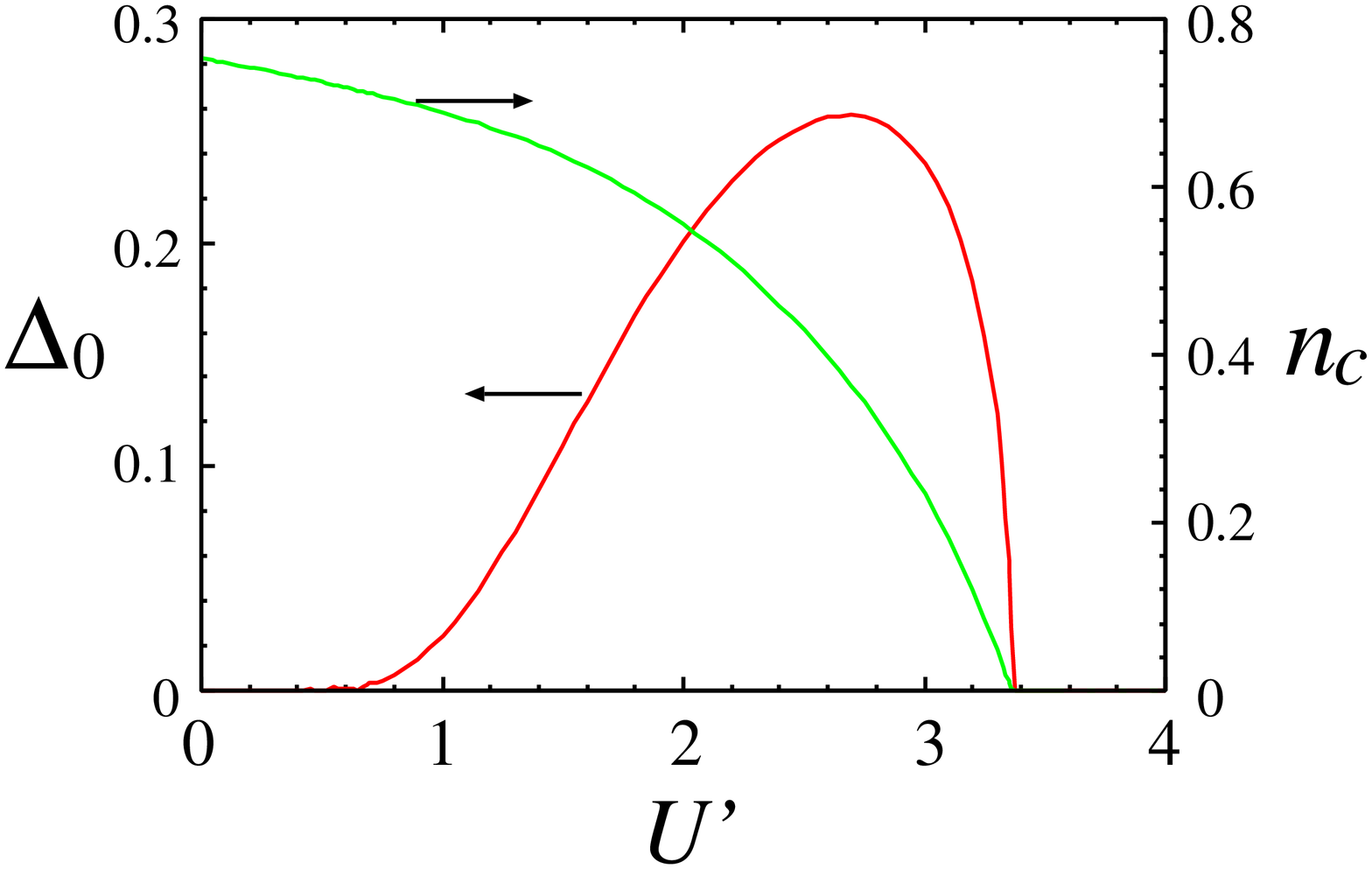}
\caption{$\Delta_0$ and $n_c$ as functions of $U^{\prime}$ with 
$U=4$ and $\mu_C=2.5$.}
\label{fig:fig10}
\end{figure}
\begin{figure}[h]
\begin{center}
\includegraphics[height=11.0cm]{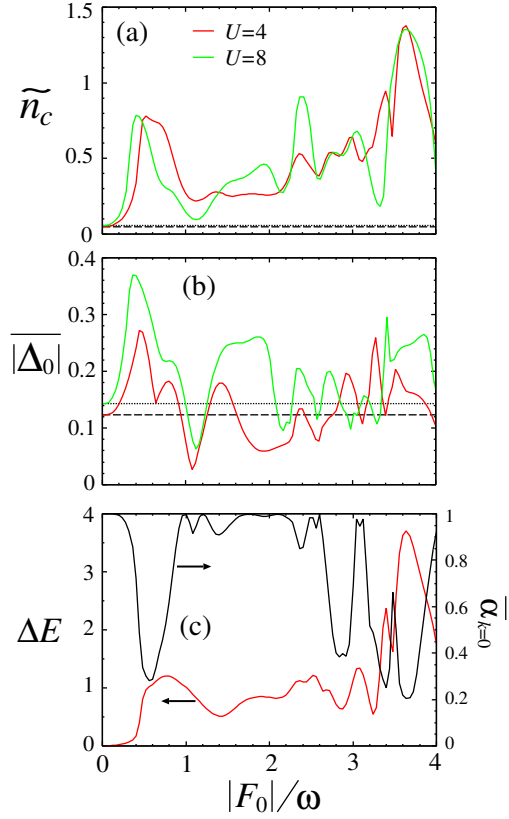}
\caption{(a) $\widetilde{n_c}$ and (b) $\overline{|\Delta_0|}$ 
as functions of $|F_0|/\omega$ for $U=4$ and $U=8$. We use 
$F(\tau)$ with the gaussian envelope. 
For $U=4$ ($U=8$), we set $U^{\prime}=3.3$ and $\mu_c=2.5$ 
($U^{\prime}=4.3$ and $\mu_c=5$). The dashed (dotted) horizontal 
lines indicate the corresponding equilibrium values for $U=4$ ($U=8$).
 (c) $\Delta E$ and $\overline{\alpha}_{k=0}$ as functions of $|F_0|/\omega$ 
for $U=4$.}
\label{fig:fig11-sin}
\end{center}
\end{figure}

\subsubsection{Ground state}
In Fig. \ref{fig:fig10}, we show $\Delta_0$ and $n_c$ 
as functions of $U^{\prime}$ in the ground state with $U=4$ and $\mu_C=2.5$. 
As in the previous studies where the Fermi surface is perfectly 
nested \cite{Kaneko_PRB12,Tanaka_PRB18}, an 
infinitesimal $U^{\prime}$ produces an EI with $\Delta_0\neq 0$. 
The order parameter $\Delta_0$ exhibits a maximum at $U^{\prime}=2.70$ 
and a transition from the EI to BI phases occurs at $U^{\prime}=U^{\prime}_{\rm cr}=3.37$ 
where $\Delta_0$ vanishes. Toward $U^{\prime}=U^{\prime}_{\rm cr}$, 
$n_c$ monotonically decreases.  In the BI phase, the $c$ and $f$ bands are 
completely decoupled so that we have $\Delta_0=0$ and $n_c=0$ ($n_f=2$). 

\subsubsection{Photoinduced dynamics}

In the calculations of photoinduced dynamics, we use two sets of 
parameters, both of which give EIs that are located near the 
EI-BI phase boundary of the ground states. 
One is $U=4$, $\mu_C=2.5$ and $U^{\prime}=3.3$. 
The other is $U=8$, $\mu_C=5$ 
and $U^{\prime}=4.3$ where $U^{\prime}_{\rm cr}=4.5$. We use $F(\tau)$ 
with the gaussian envelope [Eq. (\ref{eq:fg_tau})] and the pump-light 
frequency is tuned to the initial gap of the EI: $\omega=1.25$ 
for $U=4$ and $\omega=1.99$ 
for $U=8$. 
We note that as far as $U^{\prime}\sim U^{\prime}_{\rm cr}$, 
our results are qualitatively unaltered even when we 
choose a BI as the initial state. 
The system size is $N=200$. In analogy with the case of the atomic limit, 
we define a $k$-dependent quantity $\alpha_k$ as follows. First, we consider the overlap 
between the one-particle 
state at time $\tau$ and that in the ground state. 
If we write the one-particle state as 
$|\psi_{k\sigma}(\tau)\rangle=u_{k\sigma}(\tau)c^{\dagger}_{k\sigma}|0\rangle+
v_{k\sigma}(\tau)f^{\dagger}_{k\sigma}|0\rangle$ with 
$|u_{k\sigma}|^2+|v_{k\sigma}|^2=1$, the overlap is written as
\begin{eqnarray}
\lefteqn{|\langle \psi_{k\sigma}(\tau=0)|\psi_{k\sigma}(\tau)\rangle|^2} \nonumber \\
&=&n_c^G(k)n_c(k)+[1-n_c^G(k)][1-n_c(k)] \nonumber \\
&+&{\Delta^G}^{\ast}(k)\Delta(k)+{\Delta^G}(k)\Delta^{\ast}(k),
\label{eq:overlap-k}
\end{eqnarray} 
where ${n_c}(k)$ and $\Delta (k)$ are the momentum distribution 
function for $c$-electrons and the pair amplitude in $k$ space, 
which are written as 
\begin{equation}
n_c(k)=\langle c^{\dagger}_{k\sigma}c_{k\sigma}\rangle=|u_{k\sigma}|^2, 
\label{eq:nc-k}
\end{equation}
and 
\begin{equation}
\Delta(k)=\langle c^{\dagger}_{k\sigma}f_{k\sigma}\rangle={u_{k\sigma}}^{\ast}v_{k\sigma}, 
\label{eq:delta-k}
\end{equation}
respectively, and $n_c^G(k)$ [$\Delta^G(k)$] is 
$n_c(k)$ [$\Delta (k)$] in the ground state. 
Then, as in Eq. (\ref{eq:alpha2}), 
we define $\alpha_k$ as 
\begin{eqnarray}
\alpha_k &=&n_c^G(k)n_c(k)+[1-n_c^G(k)][1-n_c(k)] \nonumber \\
&+&2|{\Delta^G}(k)||\Delta(k)|, 
\label{eq:alpha-k}
\end{eqnarray} 
which is the upper limit of the overlap in Eq. (\ref{eq:overlap-k}). 

After the photoexcitation, $n_c$ is 
conserved and it is denoted by $\widetilde{n_c}$, whereas the time profile of $|\Delta_0|$ exhibits an oscillation 
corresponding to the Higgs amplitude mode \cite{Tanaka_PRB18}. 
The time average of $|\Delta_0|$ is denoted by $\overline{|\Delta_0|}$, 
which is defined in Eq. (\ref{eq:time_av}). 
In Fig. \ref{fig:fig11-sin}, we show $\widetilde{n_c}$, $\overline{|\Delta_0|}$, 
$\Delta E$, and $\overline{\alpha}_{k=0}$ (denoting the time average of 
$\alpha_{k=0}$) 
as functions of $|F_0|/\omega$, where $k=0$ is the location of 
the gap in the ground state. The time average is 
taken with $\tau_i=20$ and $\tau_f=50$. 
For $|F_0|/\omega\lesssim 1$, the $|F_0|/\omega$ dependence of these 
quantities is similar to that in the atomic limit shown in Fig. \ref{fig:fig3}, 
indicating that the dynamics are qualitatively 
described by the Rabi oscillation even when the bands are formed. 
As shown in Fig. \ref{fig:fig11-sin}(a), $\widetilde{n_c}$ has a peak 
at $|F_0|/\omega\simeq 0.5$ which is comparable to
 the case of the atomic limit [Fig. \ref{fig:fig3}(a)]. 
For large $|F_0|$ ($|F_0|/\omega\gtrsim 1$), a cyclic behavior of physical
 quantities that characterizes the Rabi oscillation becomes less clear. 
When we employ the rectangular envelope for $F(\tau)$, the cyclic behavior appears 
even in the region of large $|F_0|/\omega$ (Fig. \ref{fig:fig11}). 

\begin{figure}
\includegraphics[height=5cm]{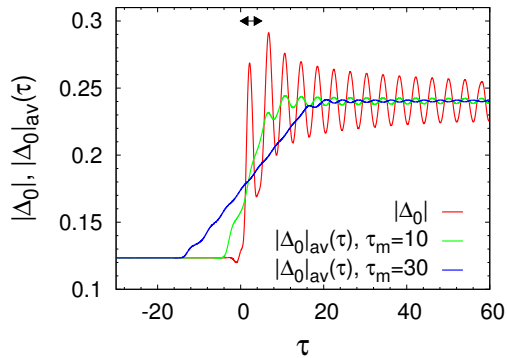}
\caption{Time profiles of $|\Delta_0|$ and $|\Delta_0|_{\rm av}(\tau)$. We use 
 $\tau_m=10$ and $30$ for $|\Delta_0|_{\rm av}(\tau)$. The other parameters are
 $U=4$, $U^{\prime}=3.3$, $\mu_C=2.5$, $\omega=1.25$, and $F_0=-0.5$. The 
double-headed arrow indicates the range $0<\tau<2\tau_w$ of application of an 
electric field.}
\label{fig:mean-order}
\end{figure}

Here we mention the choice of the values of 
$\tau_i$ and $\tau_f$ in Eq. (\ref{eq:time_av}). After the photoexcitations, physical quantities 
generally show oscillations in time. 
In the time-dependent HF method, 
the center of such an oscillation is almost constant because dephasing processes 
via electron correlations are not taken into account. 
In Fig. \ref{fig:mean-order}, we show the time profile of $|\Delta_0|$ where we 
use the $U=4$, $U^{\prime}=3.3$, $\mu_C=2.5$, $\omega=1.25$, 
and $F_0=-0.5$, which gives a large enhancement of $|\Delta_0|$. After the 
photoexcitation, $|\Delta_0|$ exhibits the 
Higgs amplitude mode with a period of about $4.0$, whereas the center of its  
oscillation is almost constant. 
In order to show explicitly how the values of $\tau_i$ and $\tau_f$ affect 
the time average, we define the time average of a physical quantity $X(\tau)$ taken 
in the range from $\tau-\tau_m/2$ to $\tau+\tau_m/2$ as,
\begin{equation}
X_{\rm av}(\tau)=\frac{1}{\tau_m}\int^{\tau+\tau_m/2}_{\tau-\tau_m/2}X(\tau^{\prime})d\tau^{\prime}.
\end{equation}
The time profiles of $|\Delta_0|_{\rm av}(\tau)$ with $\tau_m = 10$
 and $30$ are shown in Fig. \ref{fig:mean-order}, which indicates that their difference 
is very small for $\tau>20$. We note that $\overline{|\Delta_0|}$ with $\tau_i = 20$ and 
$\tau_f = 50$ presented in Fig. \ref{fig:fig11-sin}(b) corresponds 
to $|\Delta_0|_{\rm av}(\tau)$ with $\tau_m = 30$ at $\tau = 35$. 
From these results, we confirm 
that, when $\tau_m = \tau_f-\tau_i$ is larger than
the oscillation period of $|\Delta_0|$ and $\tau_i$ is taken sufficiently after the photoexcitation, 
the value of $\tau_m$ has little effects on the results. 
For the relevance to experiments, if we use $t_f=0.4$ eV for Ta$_2$NiSe$_5$ \cite{Seki_PRB14}, 
$\tau_f-\tau_i=30$ corresponds to 50 fs, which 
is comparable to time resolution of recent pump-probe measurements \cite{Mor_PRL17}. 
When $|\Delta_0|$ is small after the photoexcitation, the period of the Higgs mode
may become long. However, in such cases the amplitude of the Higgs mode 
becomes small and thus 
the choice of $\tau_i$ and $\tau_f$ does not largely affect the results.

\subsubsection{Signature of Rabi oscillation in one-particle states}

\begin{figure}[h]
\includegraphics[height=11.0cm]{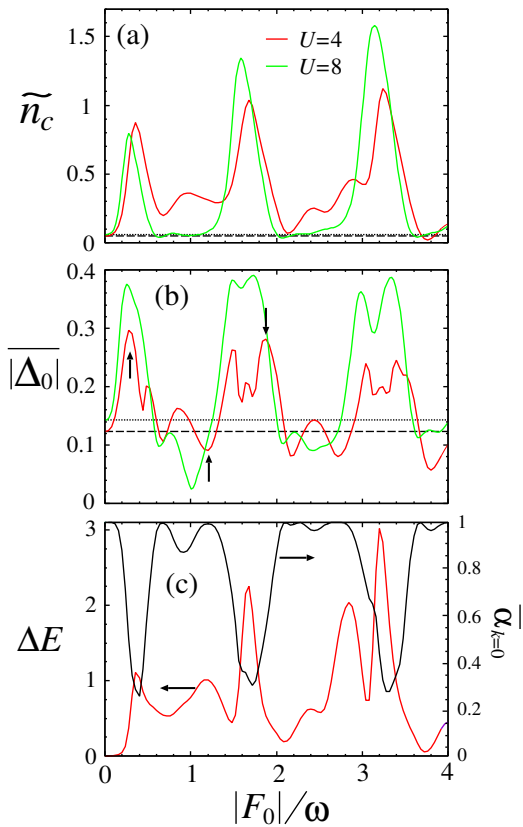}
\caption{Same plots as Fig. \ref{fig:fig11-sin} except that we use $F(\tau)$ with the  
rectangular envelope. In (b), the positions of 
$\overline{|\Delta_0|}$ with 
$|F_0|/\omega=0.3$, $1.2$, and $1.9$ for $U=4$ are indicated by 
the arrows.}
\label{fig:fig11}
\end{figure}
\begin{figure}
\includegraphics[height=4.5cm]{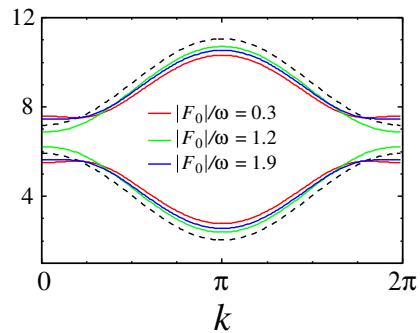}
\caption{$\overline{E}_{k\gamma\sigma}$ for different values of $|F_0|/\omega$. 
The dashed lines show the energy levels in the ground state. 
The parameters are $U=4$, $\mu_C=2.5$, and $U^{\prime}=3.3$.}
\label{fig:fig12}
\end{figure}

Here we consider $F(\tau)$ with the rectangular envelope 
[Eq. (\ref{eq:fs_tau})]. As in the case of the atomic limit, we 
discuss our results from the viewpoint of the Rabi oscillation. In Fig. \ref{fig:fig11}, we show  
$\widetilde{n_c}$, $\overline{|\Delta_0|}$, $\Delta E$, and $\overline{\alpha}_{k=0}$
 as functions of $|F_0|/\omega$. 
The oscillatory behavior
in these quantities is more evident than that in Fig. \ref{fig:fig11-sin} 
where the gaussian envelope is employed for $F(\tau)$. General tendencies of 
Figs. \ref{fig:fig11}(a), \ref{fig:fig11}(b), and \ref{fig:fig11}(c) 
are similar to those of Figs. \ref{fig:fig3-sin}(a), 
\ref{fig:fig3-sin}(b), and \ref{fig:fig3-sin}(c), respectively. 
This means that if we apply Eq. (\ref{eq:nc-rabi}) to the 
case of nonzero transfer integrals, the value of $p$ is almost 
unchanged from that in the atomic limit. 
Compared to the results with $U=4$, the oscillatory behavior is 
more prominent for those with $U=8$ where the system is closer to 
the atomic limit. 

In Fig. \ref{fig:fig12}, the time averages of the transient 
energy levels, $\overline{E}_{k\gamma\sigma}$, with $\gamma$ being 
the band index, are shown for $U=4$. The transient energy levels 
$E_{k\gamma\sigma}(\tau)$ are obtained by diagonalizing Eq. (\ref{eq:ham_mat}). 
We use $|F_0|/\omega=0.3$, $1.2$, 
and $1.9$, for which the values of $\overline{|\Delta_0|}$ are 
indicated by the arrows in Fig. \ref{fig:fig11}(b). 
For $|F_0|/\omega=0.3$ and $1.9$, the gap in $\overline{E}_{k\gamma\sigma}$ 
is larger than that in the ground state because of the enhancement of 
$\overline{|\Delta_0|}$, whereas 
it becomes smaller for $|F_0|/\omega=1.2$ where $\overline{|\Delta_0|}$ 
is suppressed. 

\begin{figure}
\includegraphics[height=5.0cm]{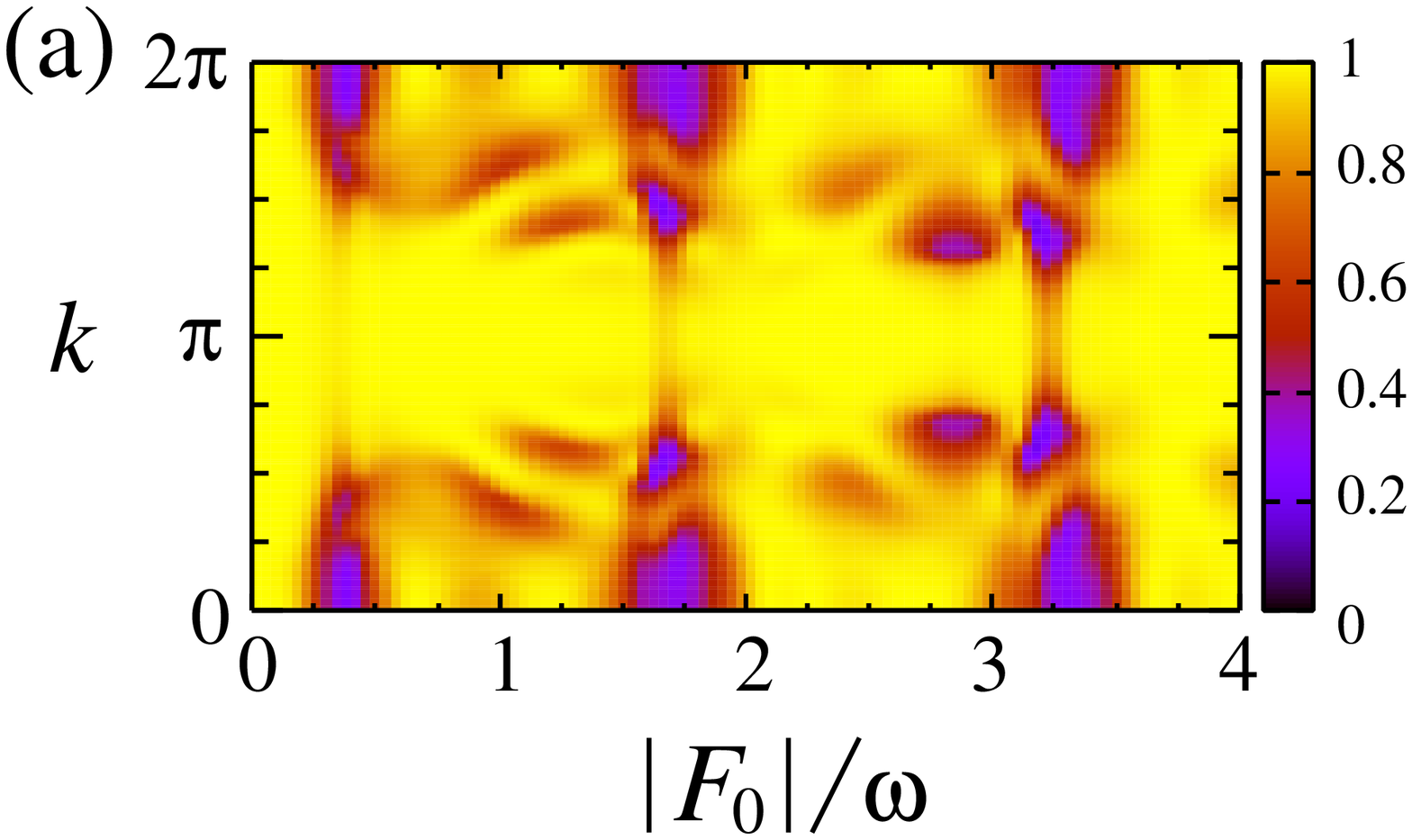}

\includegraphics[height=4.0cm]{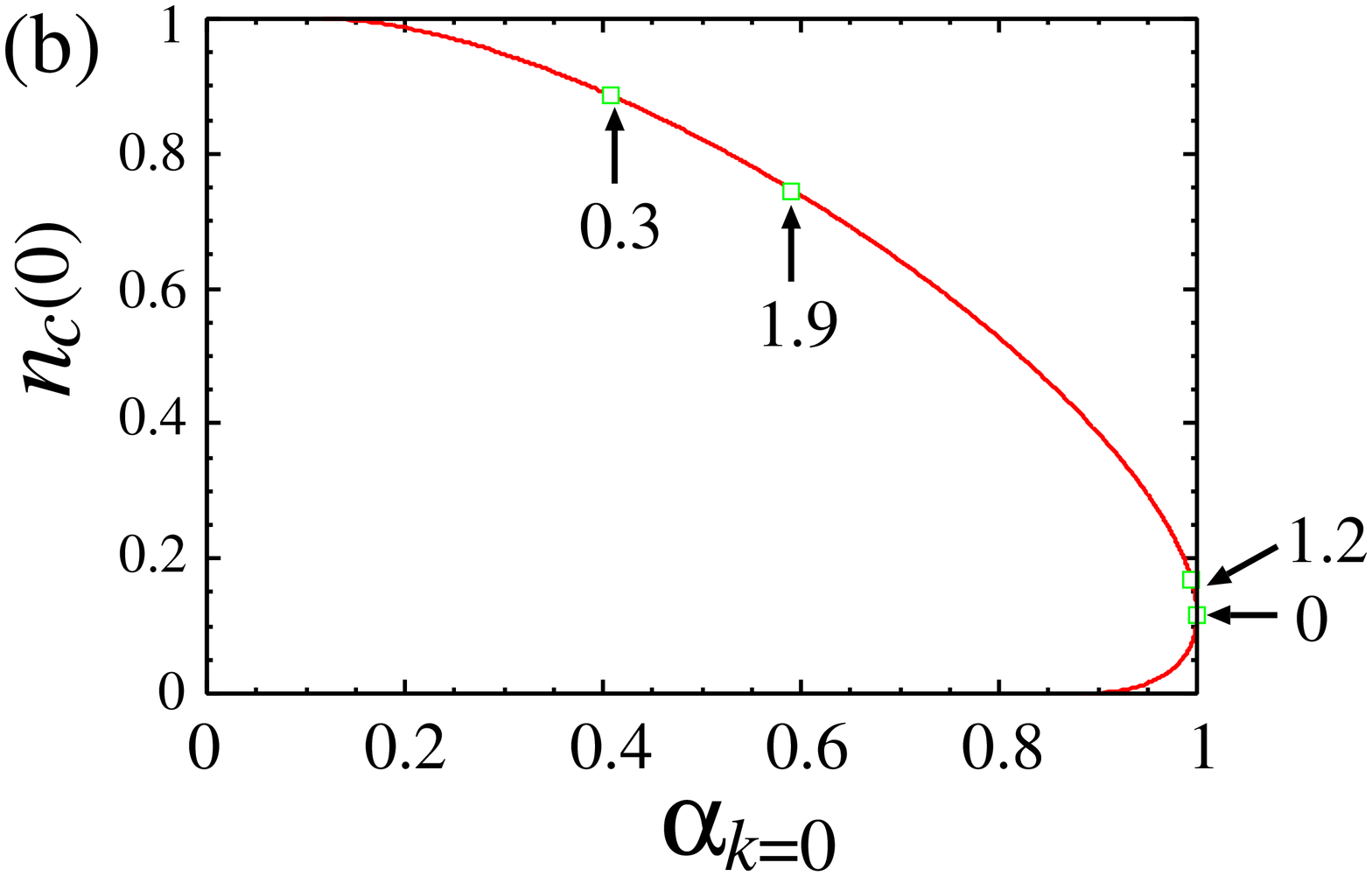}
\caption{(a) $\overline{\alpha}_k$ on $(|F_0|/\omega, k)$ plane. (b) Relation 
between $\alpha_{k=0}$ and $n_c(0)$. The position ($\alpha_{k=0}, n_c(0)$) in the 
ground state and ($\overline{\alpha}_{k=0}, \overline{n_c(0)}$) with 
$|F_0|/\omega=0.3$, $1.2$, and $1.9$ are indicated by the 
arrows. We use $U=4$, $\mu_C=2.5$, 
and $U^{\prime}=3.3$.}
\label{fig:fig13}
\end{figure}

Since the one-particle Hamiltonian is described by a $2\times2$ 
matrix, we can expect that the Rabi oscillation 
occurs for each $k$. In order to confirm this, in Fig. \ref{fig:fig13}(a) 
we show $\overline{\alpha}_k$ on the $(|F_0|/\omega,k)$ plane for $U=4$. 
It is apparent that 
$\overline{\alpha}_k$ exhibits an oscillation with respect to 
$|F_0|/\omega$. 
The oscillation amplitude depends on $k$ and is large around 
$k=0$ that is the location of the initial gap, whereas 
the period of oscillation is nearly independent of $k$. 
As shown in Fig. \ref{fig:fig11}, the periodic behavior of 
$\overline{\alpha}_{k=0}$ with respect to $|F_0|/\omega$ corresponds 
to those in $\widetilde{n_c}$, $\overline{|\Delta_0|}$, and $\Delta E$. 
By using Eqs. (\ref{eq:nc-k}) and (\ref{eq:delta-k}), we have 
\begin{equation}
|\Delta(k)|=\sqrt{n_c(k)(1-n_c(k))},
\label{eq:delta-nck}
\end{equation}
from which the relation between $\alpha_k$ and $n_c(k)$ is obtained, 
as shown in Fig. \ref{fig:fig13}(b) for the case of $k=0$. 
A similar relation is obtained even if we choose another $k$ 
(not shown). In the figure, we depict ($\alpha_{k=0}, n_c(0)$) 
in the ground state and 
($\overline{\alpha}_{k=0}, \overline{n_c(0)}$) for 
$|F_0|/\omega=0.3$, $1.2$, and $1.9$. The periodic change in the 
position as a function of $|F_0|/\omega$ is similar to that  
in the atomic limit discussed in Sect. III A. 
These results show that the periodic behavior of 
$\overline{\alpha}_k$ brings about that of $\overline{n_c(k)}$. 
Thus, the $|F_0|/\omega$ dependence of physical quantities is 
essentially caused by the Rabi oscillation of each one-particle state. 

\begin{figure}
\includegraphics[height=11.0cm]{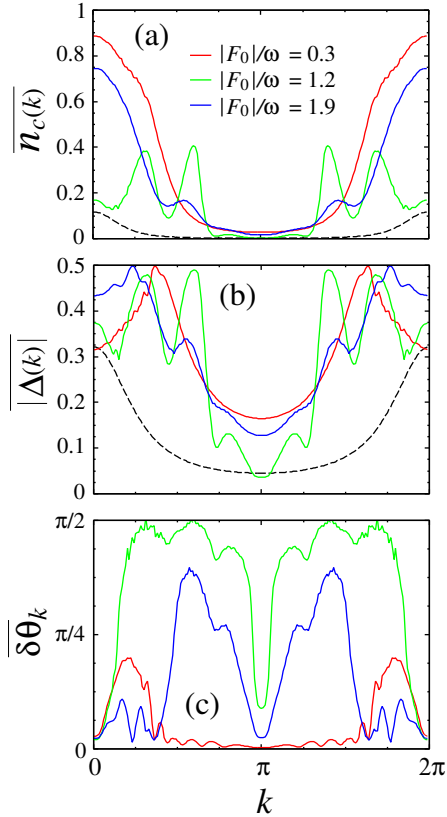}
\caption{(a) $\overline{n_c(k)}$, (b) $\overline{|\Delta(k)|}$, 
and (c) $\overline{\delta\theta}_k$ for different values of 
$|F_0|/\omega$. The dashed lines are the corresponding 
equilibrium values. 
The parameters are $U=4$, $\mu_C=2.5$, and $U^{\prime}=3.3$.}
\label{fig:fig14}
\end{figure}

In order to understand the $|F_0|/\omega$ dependence of 
$\overline{|\Delta_0|}$ in Fig. \ref{fig:fig11}(b) more accurately, 
it is necessary to discuss the phase of $\Delta(k)$ as well as 
the $k$ dependence of $n_c(k)$ and $\Delta(k)$. They have been shown 
to have an important role in determining whether the photoinduced 
enhancement of the excitonic gap occurs \cite{Tanaka_PRB18}. 
The order parameter $\Delta_0$ is related with $\Delta(k)$ by 
\begin{equation}
\Delta_0=\frac{1}{N}\sum_k \Delta(k),
\end{equation}
and we define their phases as 
\begin{equation}
\Delta(k)=|\Delta(k)|e^{i\theta_k}, 
\end{equation}
and 
\begin{equation}
\Delta_0=|\Delta_0|e^{i\theta}. 
\label{eq:d0-phase}
\end{equation}
In Fig. \ref{fig:fig14}, we show $\overline{n_c(k)}$, 
$\overline{|\Delta(k)|}$, and $\overline{\delta \theta}_k$ for 
$|F_0|/\omega=0.3$, $1.2$, and $1.9$, where 
$\delta \theta_k$ is defined as 
\begin{equation}
\delta \theta_k = 
\left\{
\begin{array}{ll}
|\theta_k-\theta| & (|\theta_k-\theta|<\pi) \\
|\theta_k-\theta|-\pi & ({\rm otherwise}).
\end{array} 
\right.
\end{equation}
In the ground state, $n_c(k)$ and $\Delta(k)$ 
have a broad $k$ dependence 
because of the BEC nature of the excitonic condensation. 
After the photoexcitation, $\overline{n_c(k)}$ and 
$\overline{|\Delta(k)|}$ are basically increased compared to their 
ground-state values. From Eq. (\ref{eq:delta-nck}), 
$|\Delta(k)|$ has its maximum value of $0.5$ when 
$n_c(k)=0.5$. When $\widetilde{n_c}$ is increased by the 
increase in $\overline{n_c(k)}$, 
the mixing between the upper and lower bands is promoted and 
$\overline{|\Delta(k)|}$ is enhanced \cite{Tanaka_PRB18}. 
However, this does not 
necessarily bring about the enhancement of $\overline{|\Delta_0|}$. 
When $|F_0|/\omega=1.2$, for instance, $\overline{|\Delta_0|}$ is smaller than 
$\Delta^G_0$ [Fig. \ref{fig:fig11}(b)] although $\widetilde{n_c}$ is 
larger than $n^G_c$ [Fig. \ref{fig:fig11}(a)]. 
As shown in Fig. \ref{fig:fig14}(c), 
$\overline{\delta \theta}_k$ is large in a 
wide region of the Brillouin zone, indicating that the enhancement 
of $\overline{|\Delta_0|}$ is hindered by the large deviation of 
$\theta_k$ from $\theta$. 
On the other hand, for $|F_0|/\omega=0.3$, $\theta_k$ is in phase 
with $\theta$ in a large area in $k$ space. 
For $|F_0|/\omega=1.9$, although $\overline{\delta \theta}_k$ 
becomes large near $k=\pm 0.6\pi$, it is small for 
$|k|\lesssim 0.4\pi$ where $\overline{|\Delta(k)|}$ has its maximum. 
Therefore, the increase in $\overline{|\Delta(k)|}$ leads to 
the enhancement of $\overline{|\Delta_0|}$. 
In short, when $\overline{|\Delta_0|}$ is enhanced by photoexcitation, 
$\theta_k$ is in phase with $\theta$ in a region where 
$\overline{|\Delta(k)|}$ is largely increased, 
whereas $\theta_k$ behaves differently from $\theta$ when 
$\overline{|\Delta_0|}$ is suppressed. In the former, 
the Rabi oscillations of one-particle states with different $k$ values 
work cooperatively to induce the gap enhancement. 

As we have shown in the previous paper \cite{Tanaka_PRB18}, when $U^{\prime}$ 
is small and the initial EI is of BCS type, the time evolution of $\theta_k$ induces a 
destructive interference to hinder the enhancement of $|\Delta_0|$, which 
is also the case for $U^{\prime}=0$. Since the excitonic 
order for small $U^{\prime}$ has a long correlation length, its 
photoinduced dynamics cannot be understood in terms of the Rabi oscillation 
in the atomic limit. However, when the initial state is of BEC type, the phases 
$\theta_k$ are nearly in phase and the Rabi oscillations for different $k$ values work 
cooperatively to enhance $|\Delta_0|$.

\section{Effects of Electron-Phonon Coupling}

We investigate effects of phonons on the photoinduced dynamics within 
the HF approximation. 
We consider the additional terms to Eq. (\ref{eq:ham}), which are used in  
\cite{Murakami_PRL17}, 
\begin{eqnarray}
\hat{H}_{\rm eph}&=&g\sum_{i\sigma}(b_i+b^{\dagger}_i)(c^{\dagger}_{i\sigma}f_{i\sigma}+f^{\dagger}_{i\sigma}c_{i\sigma}),\\
\hat{H}_{\rm p}&=&\omega_p\sum_i b^{\dagger}_ib_i,
\end{eqnarray}
where $b_i$ ($b^{\dagger}_i$) is the annihilation (creation) operator for the phonon at the 
$i$th site. 
The e-ph coupling constant and the phonon frequency are denoted by $g$ and $\omega_p$, 
respectively. We define the expectation value of the lattice displacement, 
$y_p=\langle b_i\rangle+\langle b^{\dagger}_i\rangle$, which is assumed to be independent 
of $i$. 
The time evolution of the system is computed as follows \cite{Tanaka_JPSJ10}. 
For phonons, we treat them as classical variables and numerically solve the equation of motion for $y_p$ that is written as
\begin{equation}
\frac{d^2y_p}{dt^2}=-\omega_p^2y_p-8g\omega_p{\rm Re }\Delta_0, 
\end{equation}
from which we have 
\begin{equation}
y_p=-\frac{8g}{\omega_p}{\rm Re}\Delta_0,
\label{eq:yp-delta}
\end{equation}
in the ground state. 
For the electronic part, we employ Eq. (\ref{eq:schr_eq}). In this section, we use 
$F(\tau)$ with the gaussian envelope [Eq. (\ref{eq:fg_tau})]. The results obtained by 
the rectangular envelope 
are given in Appendix C.

\subsection{Atomic limit}

\begin{figure}
\includegraphics[height=5cm]{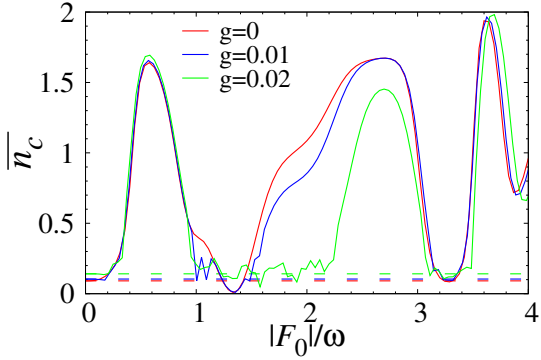}
\caption{$\overline{n_c}$ as a function of $|F_0|/\omega$ for different values of $g$ 
with $U=1$, $U^{\prime}=0.45$ and $\mu_C=0.5$ in the atomic limit. We use $\omega=0.45$, $0.458$, and 
$0.482$ for $g=0$, $0.01$, and $0.02$, respectively. 
The horizontal dashed lines indicate the corresponding equilibrium values.}
\label{fig:fig_C1}
\end{figure}

First, we discuss the case of the atomic limit ($t_f=t_c=0$). 
In the ground state, we can show that Eq. (\ref{eq:delta-nc}) 
holds even in the presence of the e-ph interaction. From Eqs. (\ref{eq:delta-nc}) and 
(\ref{eq:yp-delta}), the ground-state energy $E^{\rm AL}_g$ is written as
\begin{equation}
E^{\rm AL}_g=\frac{1}{2}(U-U^{\prime}_g)n_c^2+(\mu_C-U+U^{\prime}_g)n_c+U,
\end{equation}
where $U^{\prime}_g=U^{\prime}-8g^2/\omega_p$. This leads to $n_c=1-\mu_C/(U-U^{\prime}_g)$ 
and the critical value of $U^{\prime}$ for the EP-DP phase boundary is given by 
$U^{\prime}_{\rm cr}=U-\mu_C+8g^2/\omega_p$. 
In Fig. \ref{fig:fig_C1}, we show the time average of $n_c$, which is denoted by $\overline{n_c}$, as a 
function of $|F_0|/\omega$ for 
$g=0.01$ and $g=0.02$ with $U=1$, $U^{\prime}=0.45$, $\mu_C=0.5$, and $\omega_p=0.1$. 
For $g=0.01$ ($g=0.02$), we have $U^{\prime}_{\rm cr}=0.508$ ($U^{\prime}_{\rm cr}=0.532$). 
The value of $\omega$ is so chosen that it corresponds to the 
energy difference between the two levels. The time average is taken with $\tau_i=100$ 
and $\tau_f=400$ considering the long time-scale of phonons, $2\pi/\omega_p$. 
\begin{figure}
\includegraphics[width=7.5cm]{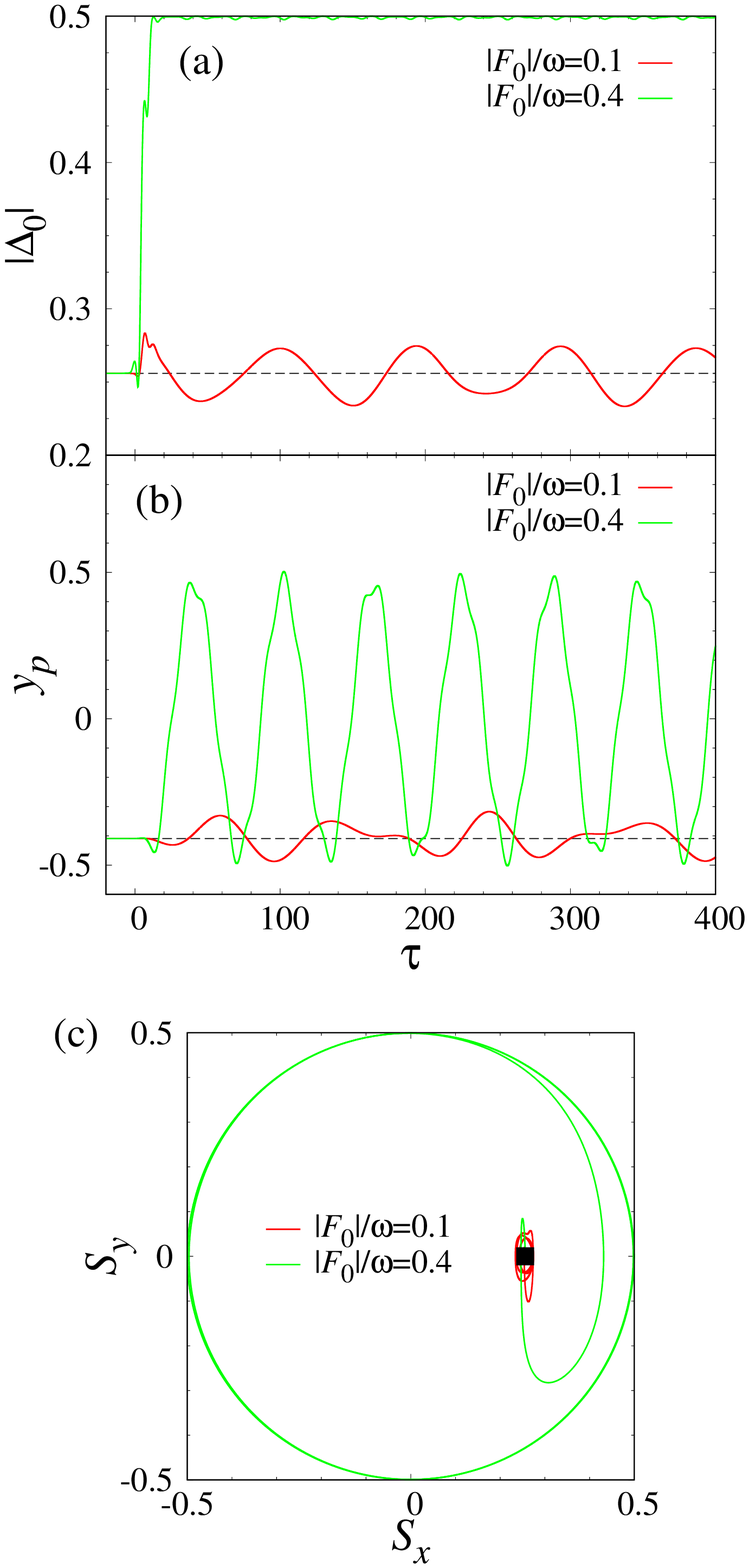}
\caption{Time evolutions of (a) $|\Delta_0|$ and (b) $y_p$, and 
(c) trajectory of $(S_x, S_y)$ for $|F_0|/\omega=0.1$ and $0.4$. 
We use $g=0.02$ and the other parameters are the same as those in Fig. \ref{fig:fig_C1}. In (a) 
and (b), the horizontal lines indicate the corresponding equilibrium values. In (c),  
we depict the time domain $-20\leq \tau \leq 300$ 
($-20\leq \tau \leq 50$) for $|F_0|/\omega=0.1$ ($|F_0|/\omega=0.4$), and 
the solid square indicates the initial position of $(S_x, S_y)$.}
\label{fig:fig_C2}
\end{figure}
\begin{figure}
\includegraphics[height=5cm]{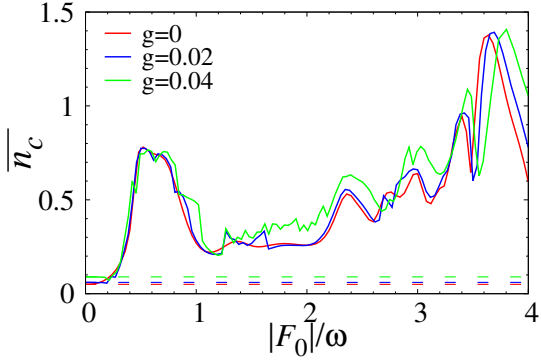}
\caption{$\overline{n_c}$ as a function of $|F_0|/\omega$ for different values of $g$ 
with $U=4$, $U^{\prime}=3.3$, and $\mu_C=2.5$ for the case of nonzero transfer 
integrals ($t_f=-t_c=1$). We use $\omega=1.25$, $1.30$, and $1.42$ for 
$g=0$, $0.02$, and $0.04$, respectively. The horizontal dashed lines indicate the corresponding 
equilibrium values.}
\label{fig:fig_C3}
\end{figure}
\begin{figure}
\includegraphics[width=7.5cm]{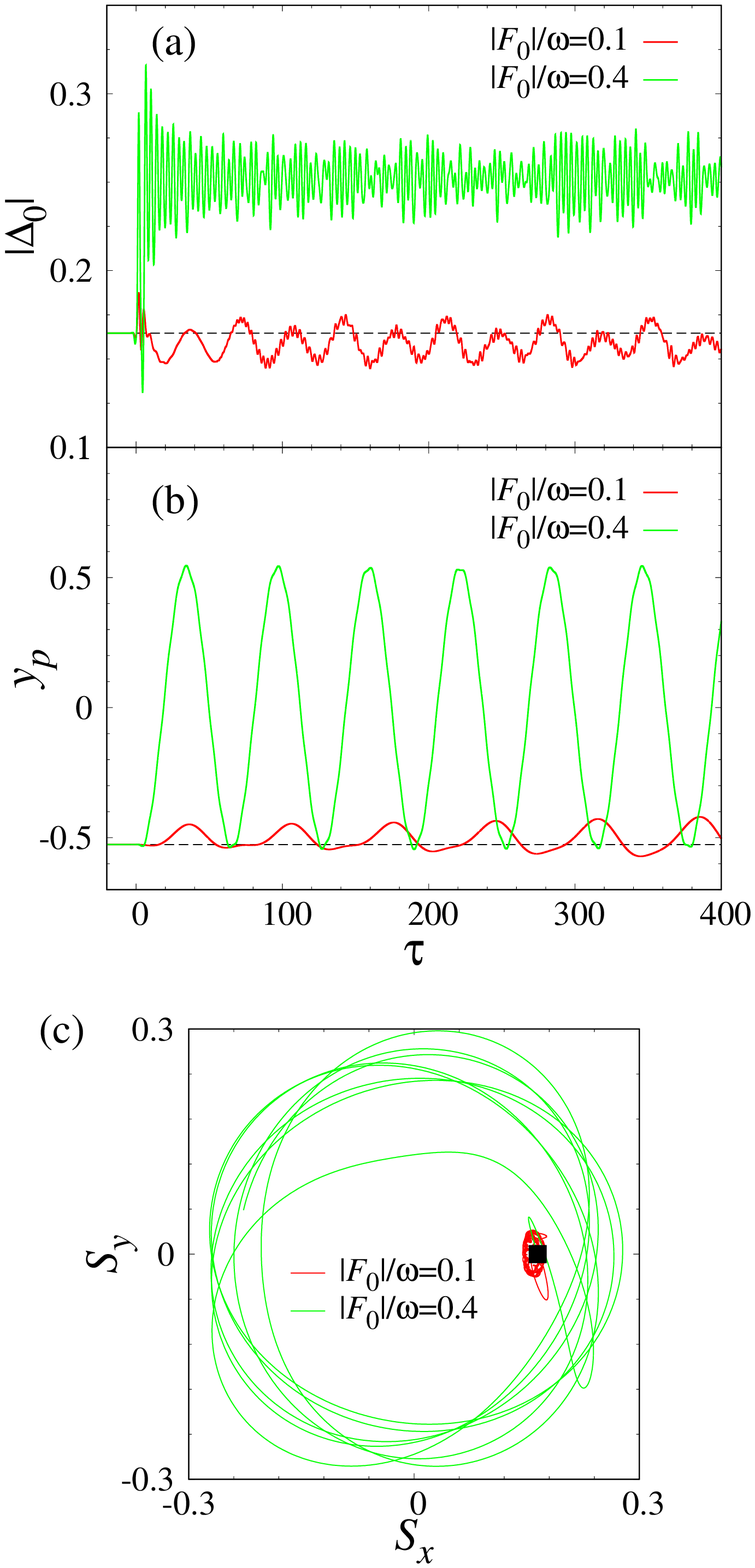}
\caption{Time evolutions of (a) $|\Delta_0|$ and (b) $y_p$, and (c) 
trajectory of $(S_x, S_y)$ for $|F_0|/\omega_0=0.1$ and $0.4$. We use $g=0.04$ 
and the other parameters are the same as those in Fig. \ref{fig:fig_C3}. In (a) 
and (b), the horizontal lines indicate the corresponding equilibrium values. In (c),  
we depict the time domain $-20\leq \tau \leq 300$ 
($-20\leq \tau \leq 50$) for $|F_0|/\omega=0.1$ ($|F_0|/\omega=0.4$), and 
the solid square indicates the initial position of $(S_x, S_y)$.}
\label{fig:fig_C4}
\end{figure}
It is apparent that the Rabi oscillation appears even with nonzero $g$. We note 
that this is also the case when we use the rectangular envelope for $F(\tau)$ 
(Appendix C). 
For $g=0.02$, we depict the time profiles of 
$|\Delta_0|$ and $y_p$ in Figs. \ref{fig:fig_C2}(a) and \ref{fig:fig_C2}(b), respectively. 
When $|F_0|$ is small ($|F_0|/\omega=0.1$), $|\Delta_0|$ and $y_p$ 
oscillate around their ground-state values. However, for large $|F_0|$ 
($|F_0|/\omega=0.4$), $|\Delta_0|$ is enhanced and $y_p$ oscillates around 
zero indicating that the effect of the lattice displacement basically disappears. 
In Fig. \ref{fig:fig_C2}(c), 
we show the trajectory of $(S_x, S_y)$ where we define $S_x={\rm Re} \Delta_0$ 
and $S_y={\rm Im} \Delta_0$ 
in the pseudospin representation. The description of the pseudospin representation 
and the trajectory of $(S_x, S_y)$ for $g=0$ are given in Appendix A. For small 
$|F_0|$, $\theta$ that is defined in Eq. (\ref{eq:d0-phase}) is confined near zero. 
This is because the 
phase mode is massive in the presence of the lattice displacement \cite{Murakami_PRL17}. 
On the other hand, $\theta$ rotates for large $|F_0|$, which is qualitatively the same as that 
for $g=0$ (Fig. \ref{fig:fig_B1} in Appendix A).

\subsection{One-dimensional model}
Next, we show results with nonzero transfer integrals ($t_f=-t_c=1$). 
We compute $\overline{n_c}$ as a function of $|F_0|/\omega$ for $g=0.02$ 
and $0.04$ where the lattice displacements in the ground state 
are $y_p=-0.217$ and $y_p=-0.527$, respectively. Here we use $U=4$, $U^{\prime}=3.3$, 
and $\mu_C=2.5$. The used value of $\omega$ corresponds to the initial gap. As 
shown in Fig. \ref{fig:fig_C3}, the introduction of the e-ph coupling does not largely affect 
the $|F_0|/\omega$ dependence of $\overline{n_c}$ as in the case of  the atomic limit. 
For $g=0.04$, we show the time profiles of 
$|\Delta_0|$ and $y_p$ in Figs. \ref{fig:fig_C4}(a) and \ref{fig:fig_C4}(b), respectively, 
whereas the trajectory of $(S_x, S_y)$ is shown in Fig. \ref{fig:fig_C4}(c). 
We use $|F_0|/\omega=0.1$ 
($\overline{|\Delta_0|}/\Delta_0(\tau=0)=0.97$) and $|F_0|/\omega=0.4$ 
($\overline{|\Delta_0|}/\Delta_0(\tau=0)=1.53$). 
The results 
are qualitatively the same as those in the atomic limit shown in Fig. \ref{fig:fig_C2}. 
These results indicate that the e-ph coupling does not have a significant role on 
the photoinduced gap enhancement based on the Rabi oscillation.

\section{Correlation effects}

In this section, we examine effects of the electron correlation that are ignored 
in the HF approximation. By using the ED method, we calculate 
ground-state properties and photoinduced dynamics of the two-orbital 
Hubbard model. We do not consider the e-ph coupling for simplicity. 
When we use single cycle pulses for photoexcitations, 
we adopt $F(\tau)$ with the gaussian envelope and 
the results obtained with the 
rectangular envelope are given in Appendix D.

\subsection{Ground state}

\begin{figure}
\includegraphics[height=5cm]{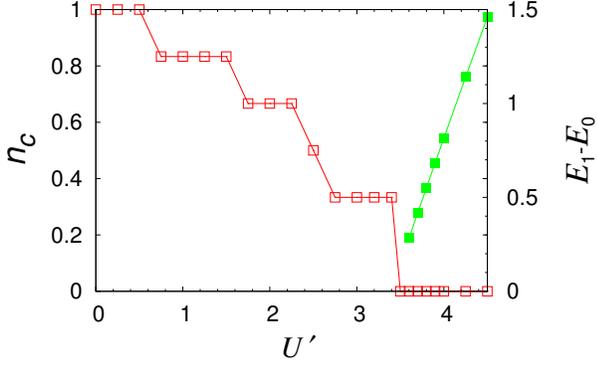}
\caption{$n_c$ as a function of $U^{\prime}$ with
 $U=4$, $\mu_C=2.5$, and $N=6$. We also show the gap $E_1-E_0$ 
for $U^{\prime}>U^{\prime}_{\rm cr}$.}
\label{fig:lanczos-gs}
\end{figure}

In the ground state, we compute the $U^{\prime}$ dependence of $n_c$ where we use 
$U = 4$, $\mu_C = 2.5$ and the system size $N=6$. As shown in Fig. \ref{fig:lanczos-gs},
 $n_c$ monotonically decreases with increasing $U^{\prime}$ and it becomes zero at 
$U_{\rm cr}^{\prime}=3.5$. This behavior is consistent with the HF
results shown in Fig. \ref{fig:fig10} where we have $U_{\rm cr}^{\prime}= 3.37$. 
The qualitative difference between the HF and ED results 
is that for $U^{\prime} < U_{\rm cr}^{\prime}$
 the excitonic order parameter $\Delta_0$ is nonzero
in the former whereas it is zero in the latter. 
We note that by the ED method we inevitably have a ground state with 
$\Delta_0= 0$ because of the finiteness of the system. 
For $U^{\prime} > U_{\rm cr}^{\prime}$, 
both methods give the BI phase with $\Delta_0= n_c = 0$ as the ground state. 
In this phase, 
the gap $E_1-E_0$ increases almost linearly with $U^{\prime}$ as shown in 
Fig. \ref{fig:lanczos-gs}, where $E_0$ and
 $E_1$ are the energies of the ground and first excited states, respectively. 
This behavior is also consistent with the HF results \cite{Tanaka_PRB18}. 
In the following,  
we consider the 
BI phase ($U^{\prime}>U^{\prime}_{\rm cr}$) as the initial state before 
photoexcitation for comparison. 

\begin{figure}
\includegraphics[height=7.5cm]{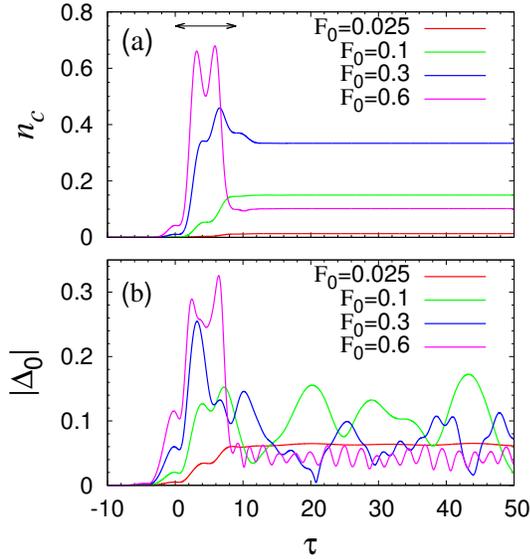}
\caption{Time evolutions of (a) $n_c$ and (b) $|\Delta_0|$ for different 
values of $F_0$ obtained by the ED method with $U=4$, $\mu_C=2.5$, 
$U^{\prime}=3.9$, $\omega=0.7$, and $N=6$. The double-headed arrow 
in (a) indicates the range $0<\tau<2\tau_w=2\pi/\omega$ of application 
of an electric field.}
\label{fig:lanczos-tp-gauss}
\end{figure}

\subsection{Photoinduced dynamics}
The time evolution of the 
system is obtained by numerically solving the time-dependent Schr$\ddot{\rm o}$dinger 
equation for the exact many-electron wave function $|\Psi(\tau)\rangle$ as
\begin{equation}
|\Psi(\tau+d\tau)\rangle =\exp\Bigl[-id\tau\hat{H}_{\rm tot}\Bigl(\tau+\frac{d\tau}{2}\Bigr)\Bigr]|\Psi(\tau)\rangle, 
\end{equation}
where $\hat{H}_{\rm tot}(\tau)=\hat{H}+\hat{H}_D(\tau)$ and we use $d\tau=0.01$.  
We use $U=4$, $\mu_C=2.5$, and  
$U^{\prime}=3.9>U^{\prime}_{\rm cr}$. 
The light frequency is set at $\omega=0.7$ 
that is near the gap $E_1-E_0=0.68$. In the following, we first show results with 
single cycle pulses and then discuss the case of CW excitations. 

\subsubsection{Excitations with single cycle pulse}
\begin{figure}
\includegraphics[height=15cm]{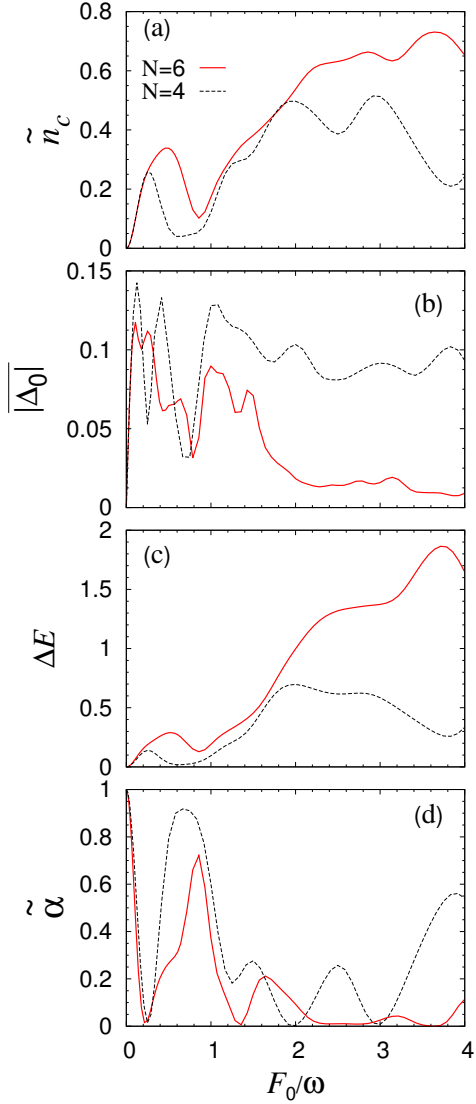}
\caption{(a) $\widetilde{n_c}$, (b) $\overline{|\Delta_0|}$, (c) $\Delta E$, and (d) 
$\widetilde{\alpha}$ as functions of $F_0/\omega$ obtained by the ED method. We use 
$F(\tau)$ with the gaussian envelope. The parameters other than $F_0$ are the same as those 
in Fig. \ref{fig:lanczos-tp-gauss}. In each panel, we show the results with $N=4$ by the dashed line 
for comparison, where $E_1-E_0=0.56$ and we use $\omega=0.6$.}
\label{fig:lanczos-time-av-gauss}
\end{figure}

\begin{figure}
\includegraphics[height=5cm]{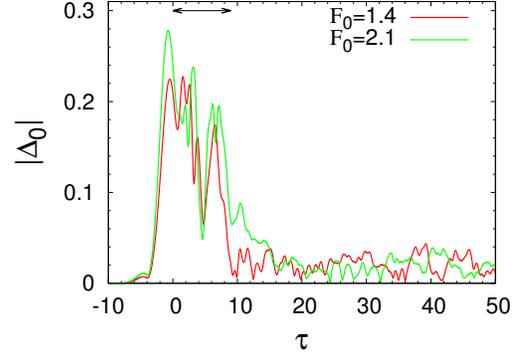}
\caption{Time evolution of $|\Delta_0|$ for 
$F_0=1.4$ and $2.1$ obtained by the ED method. The other parameters are the same as those in 
Fig. \ref{fig:lanczos-tp-gauss}. The double-headed arrow 
indicates the range $0<\tau<2\tau_w=2\pi/\omega$ of application 
of an electric field.}
\label{fig:lanczos-time-profile-gauss-large}
\end{figure}

In Fig. \ref{fig:lanczos-tp-gauss}, we show the 
time profiles of $n_c$ and $|\Delta_0|$ for different values of $F_0$ with $0<F_0/\omega\lesssim 1$. 
After the 
photoexcitation, $n_c$ is conserved, whereas $|\Delta_0|$ keeps oscillating. 
The value of $\widetilde{n_c}$ increases with increasing $F_0$, and 
then it decreases when we increase $F_0$ further ($F_0=0.6$). 
As shown in Fig. \ref{fig:lanczos-tp-gauss}(b), there is 
no clear indication of a strong dephasing in the order parameter that should suppress
 $|\Delta_0|$ after the photoexcitation with $0<F_0/\omega\lesssim 1$. 
Moreover, we do not find rapid thermalization: the oscillation 
in $|\Delta_0|$ persists long after the photoexcitation with $0<F_0/\omega\lesssim 1$. 
Although the finite size effects may play a role, our results 
at this stage do not indicate that the correlation effects seriously hinder the enhancement 
of $|\Delta_0|$.
We depict 
$\widetilde{n_c}$, $\overline{|\Delta_0|}$, $\Delta E$ and $\widetilde{\alpha}$ as functions of 
$F_0/\omega$ in Fig. \ref{fig:lanczos-time-av-gauss} where the time average of $|\Delta_0|$ 
is taken with 
$\tau_i=50$ and $\tau_f=100$. Here the overlap $\alpha$ is defined by 
$\alpha=|\langle \Psi(\tau)|\Psi(0)\rangle|^2$. After the photoexcitation, $\alpha$ is 
conserved and its value is denoted by $\widetilde{\alpha}$. 
Notably, our results indicate that for $F_0/\omega\lesssim 1$, the $F_0/\omega$ dependence 
of these 
quantities is consistent with that obtained by the HF method shown in Fig. \ref{fig:fig11-sin}. 
This strongly suggests that the enhancement of $|\Delta_0|$ as well as its interpretation 
with the help of the Rabi oscillation are robust against the correlation effects. 
We note that although the ED calculations are limited to small 
system sizes, the results with 
$N=4$ and 6 are consistent with each other. 
For $F_0/\omega\gtrsim 1$, the feature of the Rabi oscillation is unclear, which is also consistent
 with the HF results. However, $\overline{|\Delta_0|}$ obtained with $N=6$ 
is suppressed for $F_0/\omega\gtrsim 2$ 
where $\Delta E$ ($\widetilde{\alpha}$) is large (small), which is qualitatively 
different from the behavior in Fig. \ref{fig:fig11-sin}. In Fig. 
\ref{fig:lanczos-time-profile-gauss-large}, we show the time profile of $|\Delta_0|$ for 
$F_0=1.4$ ($F_0/\omega=2.0$) and $F_0=2.1$ ($F_0/\omega=3.0$). 
The value of $|\Delta_0|$ is abruptly increased by the pump light, and then it is 
 rapidly suppressed within the duration of photoexcitation. It behaves as if 
 several oscillation modes with different frequencies and phases are excited. These 
features indicate that the dephasing occurs within the duration of 
photoexcitation and it brings about the fast decay of $|\Delta_0|$. 
Although the finite size effects are expected to be substantial for 
large $F_0/\omega$, our results with $N=6$ suggest that in this region the dephasing 
has an important role in determining the value of $\overline{|\Delta_0|}$. 
This is in contrast to the case with $F_0/\omega\lesssim 1$
where $\overline{|\Delta_0|}$ can be largely enhanced. 
When we use the rectangular envelope for $F(\tau)$ [Eq. (\ref{eq:fs_tau})],  
the cyclic behavior in physical quantities becomes more evident, which we 
show in Appendix D. 

When $U^{\prime}<U^{\prime}_{\rm cr}$, our ED results 
  do not show a clear evidence of the Rabi oscillation. 
  Specifically, for 
  $U^{\prime}=3.2<U^{\prime}_{\rm cr}$, the oscillatory dependence of $\widetilde{\alpha}$ 
  and $\Delta E$ on $F_0/\omega$ that appears in Fig. 23 for the case of 
  $U^{\prime}=3.9>U^{\prime}_{\rm cr}$ 
  ($F_0/\omega\lesssim 1$) is less pronounced. 
  Although the $F_0/\omega$ dependence of
   $\overline{|\Delta_0|}$ is similar to that in Fig. 23, for $\widetilde{n_c}$ 
  the finite size effect is more severe than that with $U^{\prime}>U^{\prime}_{\rm cr}$ and
   the result with $N=4$ 
  is qualitatively different from that with $N=6$ even for $F_0/\omega<1$. 
  We speculate that these results are due to the metallic ground state with 
 $\Delta_0=0$ in the ED method.  When the system is metallic ($\Delta_0=0$), 
 it has basically gapless excitations and thus 
  it is far from a two-level system. 
  

\subsubsection{Excitations with continuous-wave laser}

\begin{figure}
\includegraphics[height=6cm]{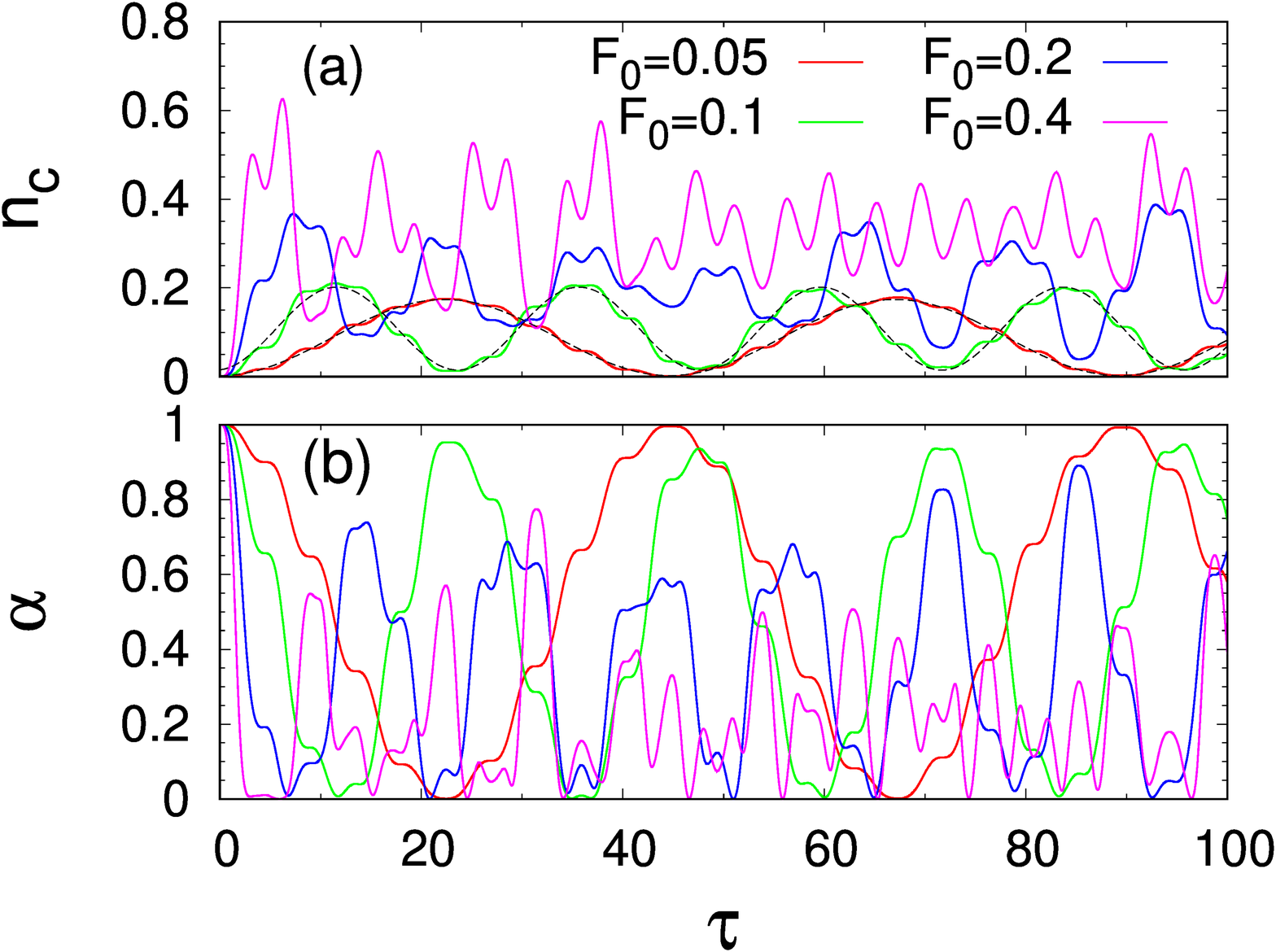}
\caption{Time evolutions of $n_c$ and $\alpha$ under CW excitations for different 
values of $F_0$ obtained by the ED method. 
We use $U=4$, $\mu_C=2.5$, $U^{\prime}=3.9$, and $\omega=0.7$. In (a), the 
dashed lines indicate the fitting results by a single sinusoidal function.}
\label{fig:nc-alpha-dynamics-CW-omg07v2}
\end{figure}

Next, we consider the case of CW excitations and examine time evolutions 
of physical quantities from the viewpoint of the Rabi oscillation. 
In Fig. \ref{fig:nc-alpha-dynamics-CW-omg07v2}, 
we show the time profiles of $n_c$ and $\alpha$ for different values of $F_0$ 
with $U^{\prime}=3.9>U^{\prime}_{\rm cr}$ for which the ground state before the 
photoexcitation is the BI. 
They exhibit an oscillation, the period of which becomes shorter 
with increasing $F_0$. 
For small $F_0$ ($F_0\lesssim 0.1$), the time profile of $n_c$ is well 
described by a single sinusoidal function of the form Eq. (\ref{eq:nc-rabi}) as shown in 
Fig. \ref{fig:nc-alpha-dynamics-CW-omg07v2}(a), and 
the minimum value in the oscillation is close to the ground-state value of $n_c$ $(=0)$. 
Correspondingly, a nearly sinusoidal oscillation appears in $\alpha$. 
It is notable that we have $\alpha\sim 1$ when $n_c\sim 0$, whereas $\alpha\sim 0$ 
when $n_c$ exhibits its maximum. 
These behaviors are consistent with the Rabi oscillation as we 
have discussed in Sect. III. With increasing $F_0$,  the oscillatory profiles in
 $n_c$ and $\alpha$ gradually become more complex. For $F_0\gtrsim 0.15$, 
 a single sinusoidal function does not fit well to the data. 
 Also, the minimum (maximum) in the oscillation of $n_c$ ($\alpha$) departs from 
 its ground-state value, which is in contrast to the case with $F_0\lesssim 0.1$. 
 
\begin{figure}
\includegraphics[height=10cm]{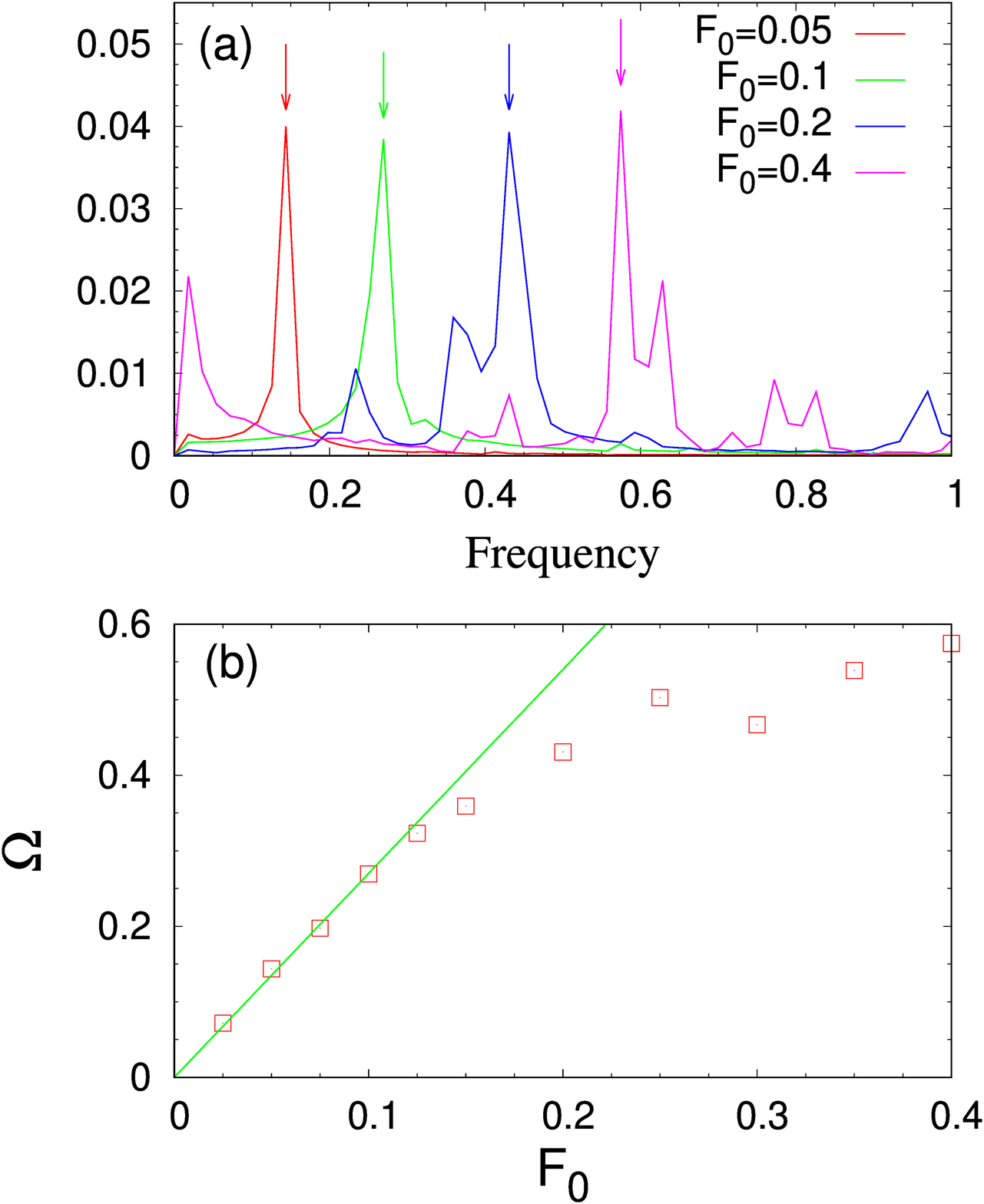}
\caption{(a) Fourier transform of $n_c$. In each spectrum, the position of 
its largest peak is indicated by the arrow. The parameters are the same 
as those of Fig. \ref{fig:nc-alpha-dynamics-CW-omg07v2}. (b) Peak frequency 
$\Omega$ in Fourier transform of $n_c$ as a function of $F_0$. The solid line is 
a fitting result to the data with $F_0\leq 0.1$.}
\label{fig:spectra-nc-CW}
\end{figure}

In Fig. \ref{fig:spectra-nc-CW}(a), we show the Fourier transform of 
$n_c$ that is calculated from the data for $50\leq \tau\leq 400$. 
There is a sharp peak in each spectrum 
and its position that is denoted by $\Omega$ becomes larger for larger $F_0$. In Fig. 
\ref{fig:spectra-nc-CW}(b), we plot the $F_0$ dependence of $\Omega$. 
For $F_0\lesssim 0.1$, $\Omega$ is nearly proportional to $F_0$: 
a function $\Omega=pF_0$ with $p=2.70$ fits well to the data. 
This result, in conjunction with the time profiles of $n_c$ and $\alpha$ shown in Fig. 
\ref{fig:nc-alpha-dynamics-CW-omg07v2}, indicates that 
for small $F_0$ the many-body dynamics under CW excitations
 is consistently interpreted from the 
viewpoint of the Rabi oscillation. 
At $F_0\sim 0.15$, $\Omega$ starts to deviate from the linear dependence on $F_0$.
At this value of $F_0$, the appearance of complex oscillatory profiles in $n_c$ and $\alpha$ 
as well as the departure of these quantities from their ground-state values 
(Fig. \ref{fig:nc-alpha-dynamics-CW-omg07v2}) are observed. These properties are different from those in the atomic limit with the 
HF approximation where the linearity characterizing the Rabi oscillation basically appears 
for large $|F_0|$ as we have discussed in Sect. III A and Appendix A. 
The deviation of the ED results with $F_0\gtrsim 0.15$ from the relation $\Omega=pF_0$ that 
is expected in two-level systems may come from effects of photoexcited electrons 
away from the gap, which should be increasingly important with increasing $F_0$. 
We note, however, that some oscillatory behavior reminiscent of the Rabi oscillation appears even for 
$F_0>0.15$, especially within the first few cycles of the 
CW excitations (Fig. 25). Therefore, in the case of single cycle pulses we can expect that 
the $F_0/\omega$ dependences of physical quantities after the photoexcitation for 
$F_0>0.15$ are qualitatively understood with the help of the Rabi oscillation. In fact, 
Fig. 23 obtained with single cycle pulses indicates the signature of the Rabi oscillation 
for $F_0/\omega\lesssim 1$ ($F_0\lesssim 0.5$). 

Finally, we examine the correspondence between the results with CW excitations and
 those with single cycle pulses in the same way as we have done in Sect. III A. We apply
  $\Omega_R^{\prime}=pF_0$ with $p=2.70$ to Eq. (\ref{eq:nc-rabi}). For $\alpha$, 
  we use Eq. (\ref{eq:alpha}) with $n^G_c=\Delta^G_0=0$. 
  By setting $\tau=2\pi/\omega$ in these equations, we can 
deduce that $\widetilde{\alpha}$ ($\widetilde{n_c}$) for single cycle pulses 
exhibits a minimum (maximum) at 
$F_0/\omega\sim 0.19$ unless the constant $A$ in Eq. (\ref{eq:nc-rabi}) 
strongly depends on $F_0$. 
 For $\widetilde{\alpha}$, this value of $F_0/\omega$ is 
 consistent with the results shown in Fig. \ref{fig:lanczos-time-av-gauss}, 
 where it exhibits a minimum at $F_0/\omega=0.21$. 
 For $\widetilde{n_c}$, its first maximum is located at $F_0/\omega=0.46$ which 
 is larger than the above estimation. This discrepancy mainly comes from an increase in 
  the amplitude of $n_c$ with increasing $F_0$ [Fig. \ref{fig:nc-alpha-dynamics-CW-omg07v2}(a)]: 
  the $F_0$ dependence of $A$ is important in determining the maximum of $n_c$. This is in contrast to 
 the time evolutions of $\alpha$ where it becomes almost zero in its first oscillation 
 irrespective of the value of $F_0$ [Fig. \ref{fig:nc-alpha-dynamics-CW-omg07v2}(b)]. 
 We note that this argument also holds for the case with the rectangular envelope 
 where the maximum of $\widetilde{n_c}$ and the minimum of $\widetilde{\alpha}$ are
  located at $F_0/\omega=0.21$ and $0.32$, respectively, as shown in Appendix D.

\section{Discussion and Summary}

Finally, we discuss possible experimental observation 
of photoinduced gap enhancement as well as the relevance of our 
results to Ta$_2$NiSe$_5$. Recent theoretical 
studies \cite{Seki_PRB14,Matsuura_JPSJ16} have shown that various equilibrium 
properties of Ta$_2$NiSe$_5$ such as the ARPES 
spectra \cite{Seki_PRB14} and the temperature dependence of magnetic 
susceptibility \cite{Salvo_JLCM86} can be reproduced by two- or three-orbital Hubbard models. 
Effects of the structural distortion observed at $T_C$ have been investigated using a 
three-orbital Hubbard model 
with e-ph interactions by the HF approximation \cite{Kaneko_PRB13}. 
It has been shown that the values of the 
e-ph interaction strengths needed to reproduce the experimentally 
observed distortion are one order of magnitude smaller than 
those of the transfer integrals and the e-e interaction strengths. Then, it has 
been argued that the EI in Ta$_2$NiSe$_5$ is ascribed to the BEC of electron-hole pairs 
which cooperatively induce the instability of 
the lattice distortion. These studies suggest that the photoinduced dynamics 
obtained in this paper based on the two-orbital Hubbard model [Eq. (\ref{eq:ham})] 
would be relevant to Ta$_2$NiSe$_5$. 

In our mechanism, photoinduced gap enhancement occurs purely electronically 
when $\omega$ is comparable to the excitonic gap. Moreover, we have shown  
that e-ph couplings do not affect our results qualitatively. 
When $\omega$ is much larger than the excitonic gap, which is the case 
in recent experiments \cite{Mor_PRL17}, a theoretical study has 
shown that e-ph couplings are crucially important for the appearance of the 
gap enhancement \cite{Murakami_PRL17}. 
Thus, our mechanism is considered as an alternative route to this phenomenon.


In this paper, we consider the case where the upper and lower bands have 
the same bandwidth ($t_c=-t_f$). However, even when the two bandwidths 
are different \cite{Kaneko_PRB13}, we expect that the gap enhancement by the Rabi 
oscillation occurs as long as the initial system is a BEC-type
EI or a nearby BI. This is because their dynamics should
be basically understood from the real-space picture \cite{Tanaka_PRB18} 
where the analysis in the atomic limit presented in this
paper is valid.


In order to examine the relevance of our results to experiments, we estimate the number of 
absorbed photons per site $n_{\rm ph}$. When $U=4$ and $U^{\prime}=3.3$ [Fig. \ref{fig:fig11-sin}(c)], 
 we have $\Delta E=0.398$ for $|F_0|/\omega=0.44$ at which 
$\overline{|\Delta_0|}$ exhibits the first peak as a function of $|F_0|/\omega$. 
This corresponds to $n_{\rm ph}=\Delta E/\omega=0.32$. We note that a 
sizable gap enhancement appears with much smaller 
values of $n_{\rm ph}$. For instance, $15$\% enhancement 
in $\overline{|\Delta_0|}$ is obtained for $|F_0|/\omega=0.2$ where 
we have $n_{\rm ph}=0.017$. In Ta$_2$NiSe$_5$, K. Okazaki {\it et al.} have 
reported that when the incident 
pump fluence is 1 ${\rm mJ}/{\rm cm}^2$, 
$n_{\rm ph}\sim 0.1$ per Ni atom whose $3d$ orbital hybridizes Se $4p$ orbital 
and forms a hole band \cite{Okazaki_NC18}. The threshold pump fluence for the appearance of 
the gap enhancement reported in Ref. 9 is $F_C=0.2$ ${\rm mJ}/{\rm cm}^2$, 
which may correspond to $n_{\rm ph}\sim 0.02$. This suggests that the pump fluence used in 
the current experimental studies is enough to observe the gap enhancement based on our 
mechanism unless $n_{\rm ph}$ depends largely on the value of the initial gap. 
However, at present a direct comparison between theoretical and experimental estimates is 
difficult by the following reasons. Firstly, in our model, we assume that the 
incident light induces the dipole transition whereas it does not 
affect the intraorbital electron motion. 
In order to realize this situation in real materials, the 
direction of light polarization as well as the crystal structure of the 
material are crucially important. For a material with a quasi-one dimensional 
structure like Ta$_2$NiSe$_5$, this indicates that the polarization of light should 
be perpendicular to the chain. The value of the matrix element for 
the dipole transition between the two bands is also important. 
Secondly, the pump-light 
frequency should be nearly tuned to the resonance condition. Note that 
in this case a recent theoretical study has shown that the gap enhancement does not 
appear when the incident light only 
affects the intraorbital electron motion \cite{Tanabe_PRB18}. 
Thirdly, the estimation of $n_{\rm ph}$ by the time-dependent 
HF method may be quantitatively inaccurate since it ignores the correlation effects 
\cite{Tanaka_JPSJ10,Miyashita_JPSJ10}. 
With regard to this point, from our ED results 
on small clusters with $U=4$ and $U^{\prime}=3.9$ where the ground state is the BI, 
we have $n_{\rm ph}=0.10$ for $F_0/\omega=0.1$ at which $\overline{|\Delta_0|}$ is 
maximally enhanced [Fig. \ref{fig:lanczos-time-av-gauss}]. This value of $n_{\rm ph}$ is 
comparable to the above-mentioned HF results. 

We note that the gap enhancement with the help of the Rabi oscillation is 
irrespective of the dimensionality of the system. In fact, 
our results for one-dimensional systems are qualitatively unaltered even in 
the two-dimensional case \cite{Tanaka_PRB18}. Moreover, our ED results suggest 
that the Rabi-oscillation-assisted gap enhancement appears even when the effects of 
quantum fluctuations are considered, although how the dephasing and thermalization 
affect the dynamics remains as a future important problem.

In summary, we investigated dynamics of EIs induced by 
electric dipole transitions using the two-orbital Hubbard model. 
Through the HF analysis of the dynamics in the atomic limit, 
we have shown that the photoinduced gap enhancement in the EI for 
single cycle pulses reported previously \cite{Tanaka_PRB18} is 
explained in terms of the Rabi oscillation. The signature of 
the Rabi oscillation appears as a periodic behavior of physical 
quantities after the photoexcitation as functions of the 
dipole field strength $F_0$. We emphasize that although 
the Rabi oscillation is a one-site problem, it represents the essential 
feature of the photoinduced dynamics in the thermodynamic limit in the 
parameter range that we have considered in this paper. We have 
performed the ED calculations 
which strongly suggest the robustness of 
this phenomenon against the correlation effects and thus corroborate 
our HF results.  The effects of the e-ph 
coupling have been examined within the HF approximation, indicating that they do 
not have a significant role on the gap enhancement in the present situation. 
Based on the present results and our previous 
work \cite{Tanaka_PRB18}, 
the condition for inducing the gap enhancement is summarized as 
follows: (i) The initial state is an EI in the BEC regime or a BI 
that is located near the EI. (ii) The pump-light frequency $\omega$ 
is near the initial gap. (iii) There is an optimal value of 
$F_0$ for enhancing the excitonic gap, which satisfies the relation 
$\Omega^{\prime}_R/\omega\sim 1/2$ with the Rabi frequency 
$\Omega^{\prime}_R\approx |F_0|$. 

\appendix

\section{Detailed dynamics in the atomic limit}

\begin{figure}
\includegraphics[height=7.5cm]{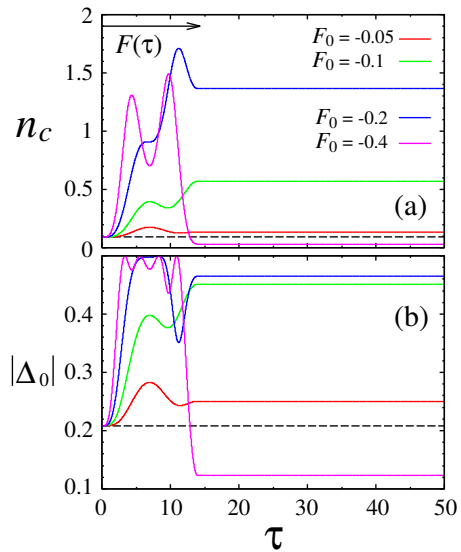}
\caption{Time evolutions of (a) $n_c$ and (b) $|\Delta_0|$ for different 
values of $F_0$ with $U=1$, $\mu_C=0.5$, $U^{\prime}=0.45$, and $\omega=0.45$. 
The arrow indicates the range where $F(\tau)$ with the rectangular envelope is nonzero. The horizontal 
dashed line in each panel indicates the corresponding equilibrium value.}
\label{fig:fig2-sin}
\end{figure}

\begin{figure}
\includegraphics[height=5.5cm]{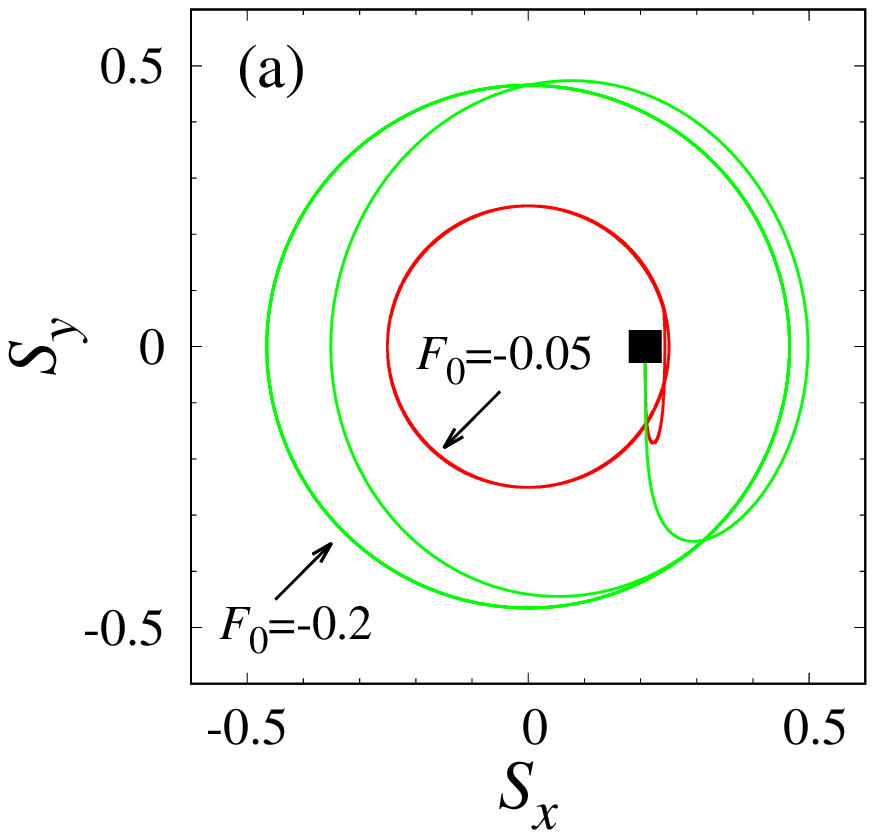}
\ \ \ \ \ \ \ \includegraphics[height=5.0cm]{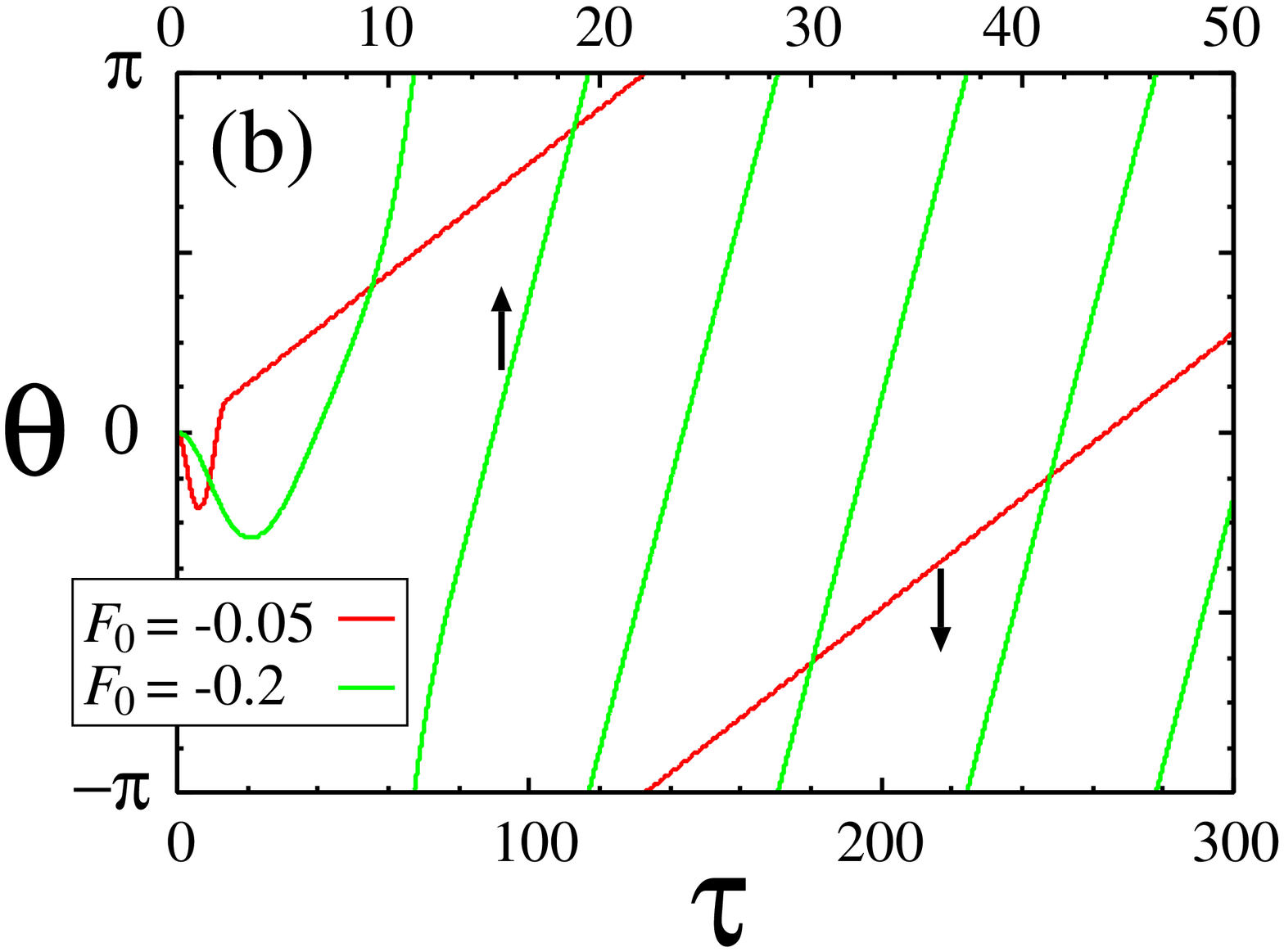}
\caption{(a) Trajectory of ($S_x, S_y$) and 
(b) time evolution of $\theta$ for $F_0=-0.05$ and $-0.2$ with 
rectangular envelope of Eq. (\ref{eq:fs_tau}). We use $U=1$, 
$\mu_C=0.5$, $U^{\prime}=0.45$ and $\omega=0.45$. 
For $F_0=-0.05$ ($F_0=-0.2$), 
we show the time domain $0\leq \tau \leq 300$ ($0\leq \tau \leq 50$). In (a), 
the solid square indicates the initial position of ($S_x, S_y$).}
\label{fig:fig_B1}
\end{figure}

We show details of the real-time dynamics of mean-field order parameters 
in the atomic limit. We use $F(\tau)$ with the rectangular envelope [Eq. (\ref{eq:fs_tau})]. 
The time profiles of $n_c$ and $\Delta_0$ for different values of $F_0$ are shown 
in Fig. \ref{fig:fig2-sin} where the parameters are the same as those in 
Fig. \ref{fig:fig2}. We introduce the psedospin operators as 
\begin{equation}
\hat{S}_{\gamma}\equiv \Psi^{\dagger}\frac{1}{2}\sigma_{\gamma}\Psi, 
\label{eq:B1}
\end{equation}
where $\sigma_{\gamma}$ ($\gamma=x, y, z$) are the Pauli matrices 
and we omit the spin index in $\Psi_{\sigma}$ for brevity. 
With this representation, the expectation values of the pseudospin 
$S_{\gamma}=\langle \hat{S}_{\gamma}\rangle$ components are written as
\begin{align}
S_{x}(\tau)&= {\rm Re}\Delta_0,\tag{A2a} \\
S_{y}(\tau)&= {\rm Im}\Delta_0,\tag{A2b}\\
S_{z}(\tau)&= \frac{1}{4}(n_c-n_f),\tag{A2c}
\label{eq:B2}
\end{align}
which give $\Delta_0=S_x+iS_y$ and $n_c=2S_z+1$. 
By using the equation of motion for the pseudospin operators, the time evolution 
of ${\bm S}=(S_x,S_y,S_z)$ is given by
\setcounter{equation}{2}
\begin{equation}
\partial_{\tau}{\bm S}={\bm B}(\tau)\times {\bm S}(\tau), 
\label{eq:B3}
\end{equation}
where 
\begin{align}
B_x&=2[U^{\prime}{\rm Re} \Delta_0-F(\tau)], \tag{A4a} \\
B_y&=2U^{\prime}{\rm Im} \Delta_0, \tag{A4b} \\
B_z&=-(\epsilon^c-\epsilon^f). \tag{A4c}
\label{eq:B4}
\end{align}
In Figs. \ref{fig:fig_B1}(a) and \ref{fig:fig_B1}(b), we show the trajectory of 
$(S_x, S_y)$ and the time evolution of $\theta$ that has been defined in 
Eq. (\ref{eq:d0-phase}), respectively, for single 
cycle pulses with $F_0=-0.05$ and $-0.2$. We use $\mu_C=0.5$, $U^{\prime}=0.45$, 
and $\omega=0.45$.  
For $F_0=-0.05$, $|\Delta_0|$ ($=\sqrt{{S_x}^2+{S_y}^2}$) is slightly increased by 
the photoexcitation, whereas it is largely enhanced for $F_0=-0.2$. 
After the photoexcitation, the value of 
$|\Delta_0|$ is conserved and $\theta$ rotates with almost a constant velocity. 
As we increase $|F_0|$, the velocity becomes larger as shown in Fig. 
\ref{fig:fig_B1}(b).

\begin{figure}
\includegraphics[height=5.0cm]{
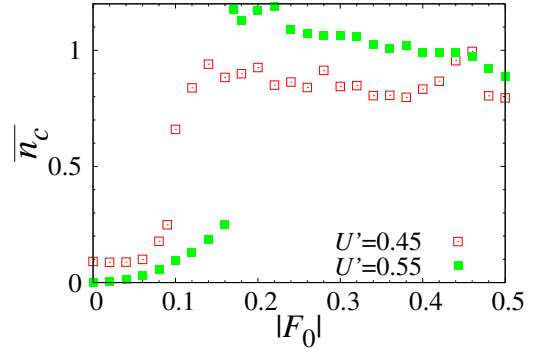}
\caption{$\overline{n_c}$ as a function of $|F_0|$ with $U^{\prime}=0.45$ 
and $0.55$ for the case of CW excitations. 
We use $\mu_C=0.5$ and $\omega=0.45$ ($\omega=0.6$) for $U^{\prime}=0.45$ ($U^{\prime}=0.55$).}
\label{fig:fig_B2}
\end{figure}

Next, we discuss results under CW excitations. As we have shown in Fig. 
\ref{fig:fig5}, $n_c$ oscillates near its ground-state value for small $|F_0|$ 
($F_0=-0.05$), 
whereas it exhibits a large oscillation for large $|F_0|$ ($F_0\lesssim -0.1$). 
Figure \ref{fig:fig_B2} 
shows $\overline{n_c}$ ($=2\overline{S_z}+1$) as a function of $|F_0|$ for 
$U^{\prime}=0.45$ and $0.55$. 
For both cases, there is a threshold $F^c_0$ at which $\overline{n_c}$ 
abruptly increases. We obtain $F^c_0=-0.1$ for $U^{\prime}=0.45$ and 
$F^c_0=-0.16$ for $U^{\prime}=0.55$. Such a dynamical transition has been 
previously reported in one-dimensional excitonic insulators 
within the HF theory \cite{Murakami_PRL17}.

\begin{figure}
\includegraphics[height=5.5cm]{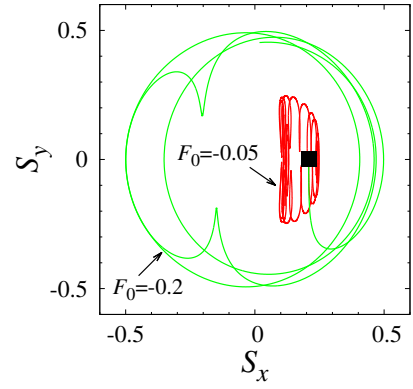}
\caption{Trajectory of ($S_x, S_y$) 
under CW excitations for $F_0=-0.05$ and $-0.2$. 
We use $U^{\prime}=0.45$ and $\omega=0.45$. 
For $F_0=-0.05$ ($F_0=-0.2$), we show the time domain $0\leq \tau \leq 300$ 
($0\leq \tau \leq 50$). The open square indicates the initial point of ($S_x, S_y$).}
\label{fig:fig_B3}
\end{figure}

\begin{figure}
\includegraphics[height=10.0cm]{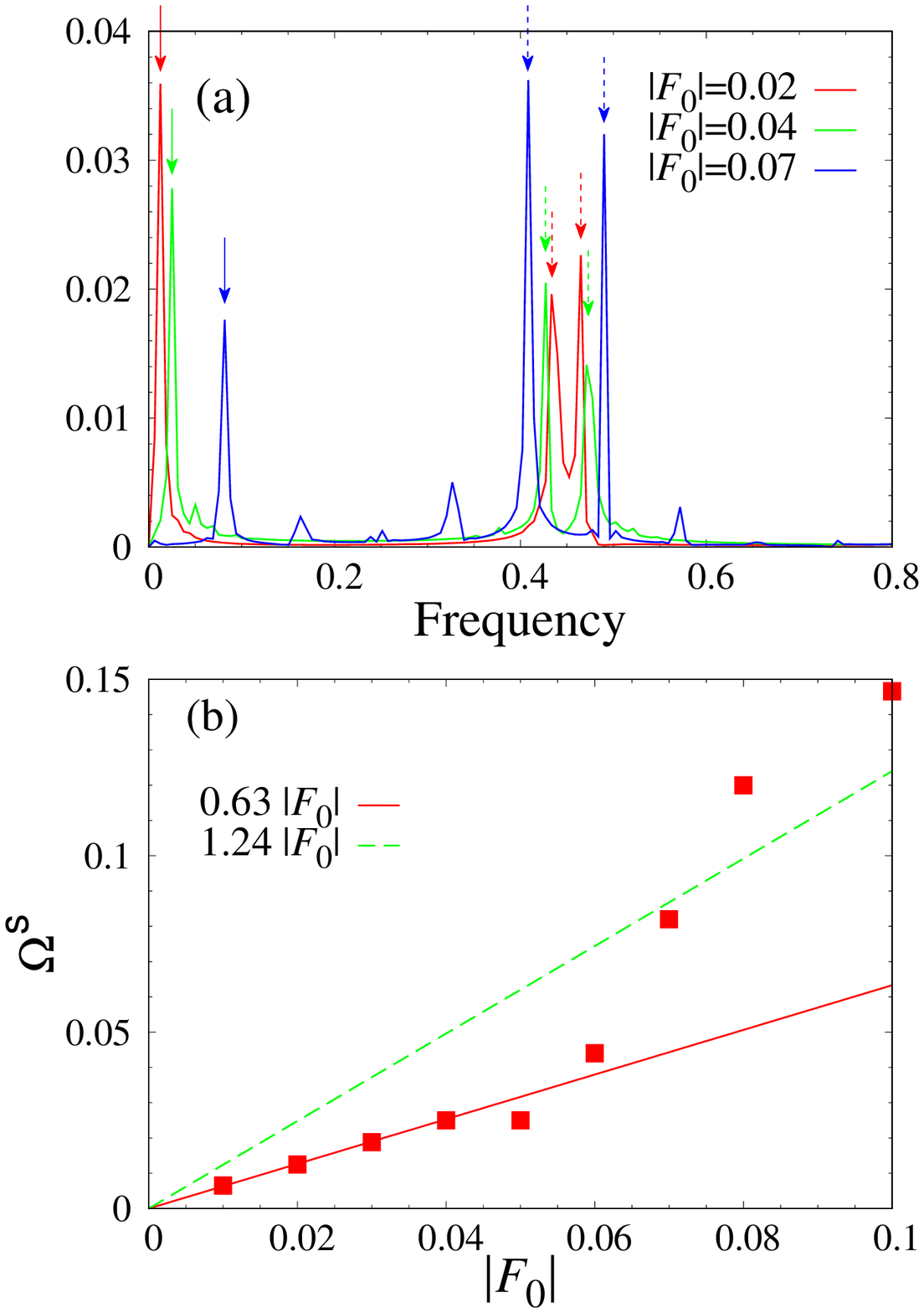}
\caption{(a) Fourier transform of $n_c$ for $|F_0|<|F^c_0|$ with 
$U^{\prime}=0.45$ and $\omega=0.45$. The peaks corresponding to $\Omega^S$ 
($\Omega^{\pm f}$) are indicated by the solid (dashed) arrows. 
(b) $|F_0|$ dependence of $\Omega^S$. The solid line is fit to 
the data with $|F_0|\leq0.06$. The fitting result in Fig. \ref{fig:fig7} for $U^{\prime}=0.45$ 
is also shown by the dashed line.}
\label{fig:fig_B4}
\end{figure}

In the following, we examine the difference between the dynamics for 
$|F_0|<|F^c_0|$ and that for $|F_0|>|F^c_0|$. 
First, we consider the case of $U^{\prime}=0.45$ where the initial state 
is in the EP. We show the trajectory of 
($S_x, S_y$) with $F_0=-0.05$ ($|F_0|<|F^c_0|$) 
and $-0.2$ ($|F_0|>|F^c_0|$) under CW excitations with $\omega=0.45$ 
in Fig. \ref{fig:fig_B3}. 
For $F_0=-0.05$, ($S_x, S_y$) is bound near the 
ground-state position, whereas it is unbound for $F_0=-0.2$. 
This corresponds to bound and unbound oscillations in $n_c$ for 
$F_0=-0.05$ and $-0.2$ (Fig. \ref{fig:fig5}), respectively. 
In Fig. \ref{fig:fig_B4}(a), we show the Fourier transform of $n_c$ for small 
$|F_0|$ ($<|F^c_0|$), indicating that $n_c$ has one slow oscillation component 
with frequency $\Omega^S\lesssim 0.15$ and two fast components with frequencies 
$\Omega^{f\pm}$ near $\omega$, which we can write as 
$\Omega^{f\pm}=\omega\pm\delta \Omega$. 
Both $\Omega^S$ and $\delta \Omega$ 
increase with increasing $|F_0|$. When $|F_0|$ is small ($|F_0|<0.06$), the 
peak at $\Omega^S$ is dominant, whereas those at $\Omega^{f\pm}$ become 
dominant for $0.06<|F_0|<|F^c_0|$. 
As we increase $|F_0|$ further ($|F_0|>|F^c_0|$), the spectra change drastically 
as we have shown in Fig. \ref{fig:fig6}. 
In Fig. \ref{fig:fig_B4}(b), we show the $|F_0|$ dependence of $\Omega^S$. 
When $|F_0|\lesssim 0.06$, $\Omega^S$ is proportional to $|F_0|$ and we have 
$\Omega^S=p|F_0|$ with $p=0.63$ that is different from the value ($p=1.24$) 
obtained in Sect. III A for $|F_0|>|F^c_0|$. The region of $|F_0|$ ($0.06<|F_0|<|F^c_0|$) where the value of $p$ 
largely deviates coincides with that where $\overline{n_c}$ 
exhibits the abrupt increase in Fig. \ref{fig:fig_B2}.
These results indicate that the dynamics for 
$|F_0|<|F^c_0|$ have a character different from that for $|F_0|>|F^c_0|$. 

\begin{figure}
\includegraphics[height=5.5cm]{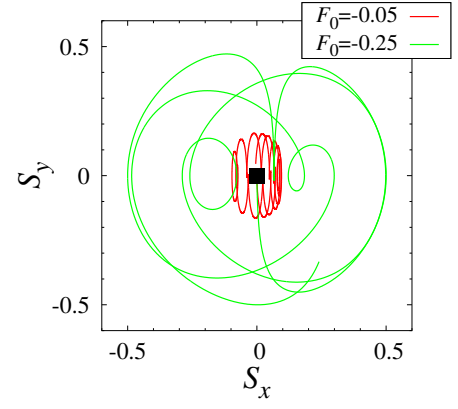}
\caption{Similar plot as Fig. \ref{fig:fig_B3} for $U^{\prime}=0.55$ 
and $\omega=0.6$. We show the time domain of $0\leq \tau \leq 150$ ($0\leq \tau \leq 50$) 
for $F_0=-0.05$ ($F_0=-0.2$).}
\label{fig:fig_B5}
\end{figure}

\begin{figure}
\includegraphics[height=10.0cm]{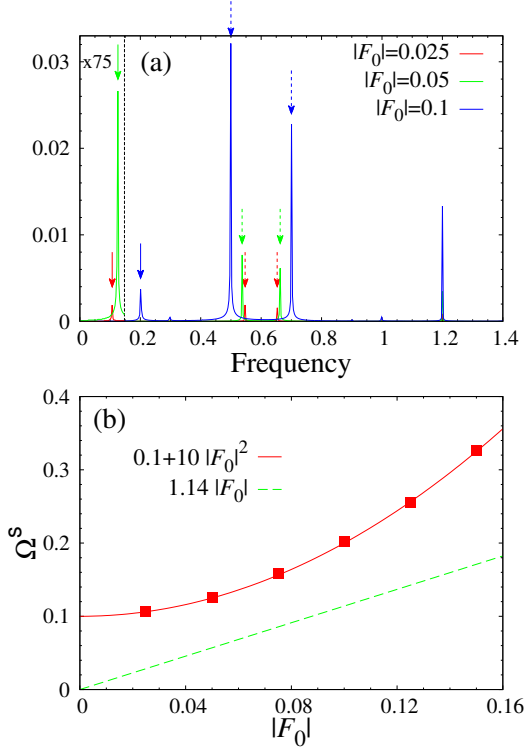}
\caption{(a) Fourier transform of $n_c$ for $|F_0|<|F^c_0|$ with 
$U^{\prime}=0.55$ and $\omega=0.6$. The peaks corresponding to $\Omega^S$ 
($\Omega^{\pm f}$) are indicated by the solid (dashed) arrows.  
(b) $|F_0|$ dependence of $\Omega^S$. The solid curve indicates the fitting result. 
The fitting result in Fig. \ref{fig:fig7} for $U^{\prime}=0.55$ is also shown by the dashed line.}
\label{fig:fig_B6}
\end{figure}

Next, we show results with $U^{\prime}=0.55$ where the initial state is 
in the DP. The trajectory of ($S_x, S_y$) 
for $F_0=-0.05$ ($|F_0|<|F^c_0|$) and 
$F_0=-0.25$ ($|F_0|>|F^c_0|$) with $\omega=0.6$ is depicted in Fig. \ref{fig:fig_B5}. 
Similar to the case of $U^{\prime}=0.45$, 
$(S_x, S_y)$ is bound near its initial position for $|F_0|<|F^c_0|$, whereas it is 
unbound for $|F_0|>|F^c_0|$. 
In Fig. \ref{fig:fig_B6}(a), we show 
the Fourier transform of $n_c$ for $|F_0|<|F^c_0|$. 
The dominant oscillation components in $n_c$ have frequencies 
$\Omega^{f\pm}= \omega\pm \delta\Omega$, and there is a slow oscillation component with 
$\Omega^S= 2\delta\Omega$ whose amplitude is higher order in $|F_0|$. 
When $|F_0|$ is small, 
we can solve Eq. (\ref{eq:B3}) in the lowest order of $F_0$ with the initial 
condition ${\bm S}(\tau=0)=(0,0,-1/2)$ as
\begin{align}
S^x(\tau)&= \frac{F_0}{\omega^2-a^2}(a\sin \omega \tau-\omega \sin a\tau),\tag{A5a} \\
S^y(\tau)&= \frac{\omega F_0}{\omega^2-a^2}(\cos a\tau-\cos \omega \tau),\tag{A5b}\\
S^z(\tau)&= \frac{\omega F_0}{\omega^2-a^2}\Bigl[\frac{1-\cos (\omega+a)\tau}{\omega+a}\nonumber \\
&+\frac{1-\cos (\omega-a)\tau}{\omega-a}+\frac{1-\cos 2\omega \tau}{2\omega}\Bigr],\tag{A5c}
\label{eq:B5}
\end{align}
where $a=\mu_C-U+U^{\prime}$. From Eq. (A5c), we find that for $|F_0|\rightarrow 0$, 
$\delta \Omega\rightarrow a=0.05$ and there is an oscillation component with frequency 
$2\omega=1.2$, which are consistent with the numerical results shown in Fig. 
\ref{fig:fig_B6}(a). 
As we increase $|F_0|$, $\Omega^S$ increases. 
$\Omega^S=0.1+10|F_0|^2$ is fit well to the data as shown in Fig. \ref{fig:fig_B6}(b). 
These results indicate that the 
dynamics for $|F_0|<|F^c_0|$ is essentially different from that for $|F_0|>|F^c_0|$ 
as in the case of $U^{\prime}=0.45$. 
In fact, the spectra of $n_c$ for $|F_0|>|F^c_0|$ (not shown) 
are largely different from those for $|F_0|<|F^c_0|$.

\section{Effects of $\tau$-dependence of $n_c$ and $\Delta_0$  
in Eq. (\ref{eq:ham_mat_al}) on the dynamics}

\begin{figure}[h]
\includegraphics[height=8.5cm]{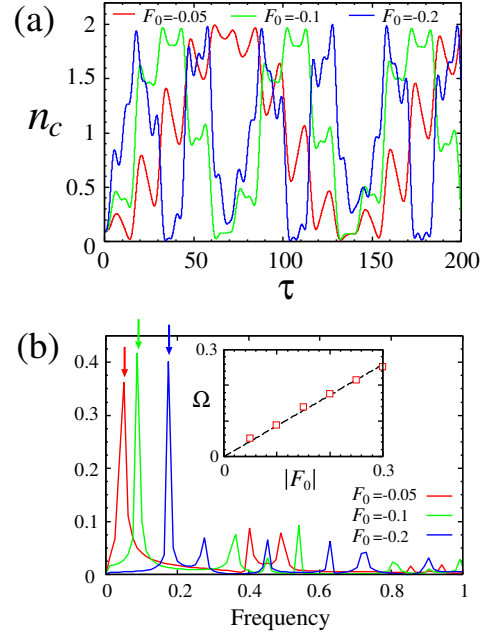}
\caption{(a) Time profile of $n_c$ under CW excitations obtained 
by the time evolution 
operator in which we artificially replace $n_c$ and $\Delta_0$ 
by those at $\tau=0$. 
(b) Fourier transform of (a). In each spectrum, the position of 
the largest peak, $\Omega$, is indicated by the arrow. In the inset, 
the $|F_0|$ dependence of $\Omega$ is shown, where the fitting result 
is also depicted by the dashed line.}
\label{fig:fig9}
\end{figure}

As we have mentioned in III A {\it 3}, Eq. (\ref{eq:ham_mat_al}) 
possesses $\tau$-dependent mean-field order 
parameters from which the time evolution 
operator is constructed. 
In order to examine how their $\tau$-dependence affects 
the dynamics in the atomic limit, we artificially replace $n_c$ and $\Delta_0$ 
in the time evolution operator by $n^G_c$ and $\Delta^G_0$, respectively, 
and compute the time profile of $n_c$ under CW excitations. 
The parameters we used are the same as those in Fig. \ref{fig:fig5}.
 The results are shown in 
Fig. \ref{fig:fig9}(a).  
Compared with Fig. \ref{fig:fig5}, a large oscillation 
in $n_c$ appears even when $|F_0|$ is small. 
From the Fourier spectra 
shown in Fig. \ref{fig:fig9}(b), we obtain $\Omega \sim p|F_0|$ with $p=0.86$. 
This result indicates that the $\tau$ dependence of the order 
parameters is important in determining the dynamics for small 
$|F_0|$ ($|F_0|<0.1$). However, it does not alter the 
dynamics qualitatively for larger $|F_0|$ 
where the Rabi oscillation appears in Fig. \ref{fig:fig5}. 
These facts give a reason why the quantitative 
difference between the value of $\widetilde{n_c}$ for single 
cycle pulses and that of $n_c$ at $\tau=2\pi/\omega$ computed from 
Eq. (\ref{eq:nc-rabi}) becomes large for small $|F_0|$ 
($|F_0|/\omega \lesssim 0.3$), which can be seen in Fig. \ref{fig:fig8}. 

\section{HF results in the presence of phonons for the case of 
rectangular-envelope pulse}

\begin{figure}
\includegraphics[height=5cm]{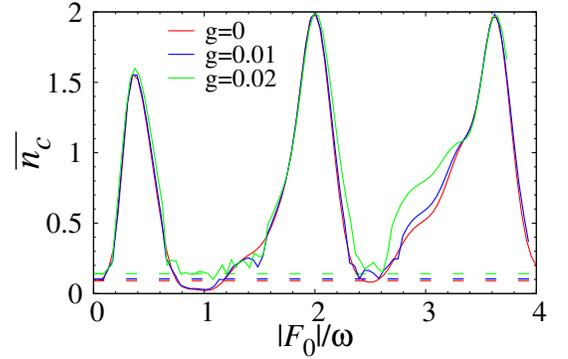}
\caption{Same plot as Fig. \ref{fig:fig_C1} except that we use $F(\tau)$ with 
the rectangular envelope.}
\label{fig:fig-phonon-atm-sin}
\end{figure}

We show  the $|F_0|/\omega$ dependence of 
$\overline{n_c}$ in the presence of the e-ph coupling when we use  
$F(\tau)$ with the rectangular envelope. 
In Fig. \ref{fig:fig-phonon-atm-sin}, the results in the atomic limit are depicted. 
The parameters are the same as those in Fig. \ref{fig:fig_C1}. 
\begin{figure}
\includegraphics[height=5cm]{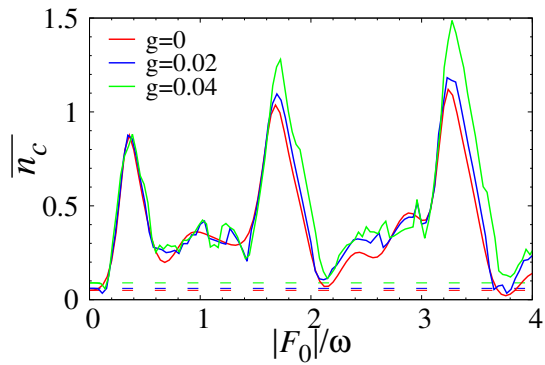}
\caption{Same plot as Fig. \ref{fig:fig_C3} except that we use $F(\tau)$ with 
the rectangular envelope.}
\label{fig:fig-phonon-tf1-sin}
\end{figure}
For the one-dimensional model with $t_f=1$ and $t_c=-1$, the results are 
shown in Fig. \ref{fig:fig-phonon-tf1-sin} 
where the parameters are the same as those in Fig. \ref{fig:fig_C3}. 
From Figs. \ref{fig:fig-phonon-atm-sin} and \ref{fig:fig-phonon-tf1-sin},
 we confirm that the e-ph coupling has little effects on the $|F_0|/\omega$ 
dependence of $\overline{n_c}$ as in the case of the gaussian-envelope 
pulse shown in Figs. \ref{fig:fig_C1} and \ref{fig:fig_C3}.

\section{ED results for the case of rectangular-envelope pulse}

\begin{figure}[h]
\includegraphics[height=15cm]{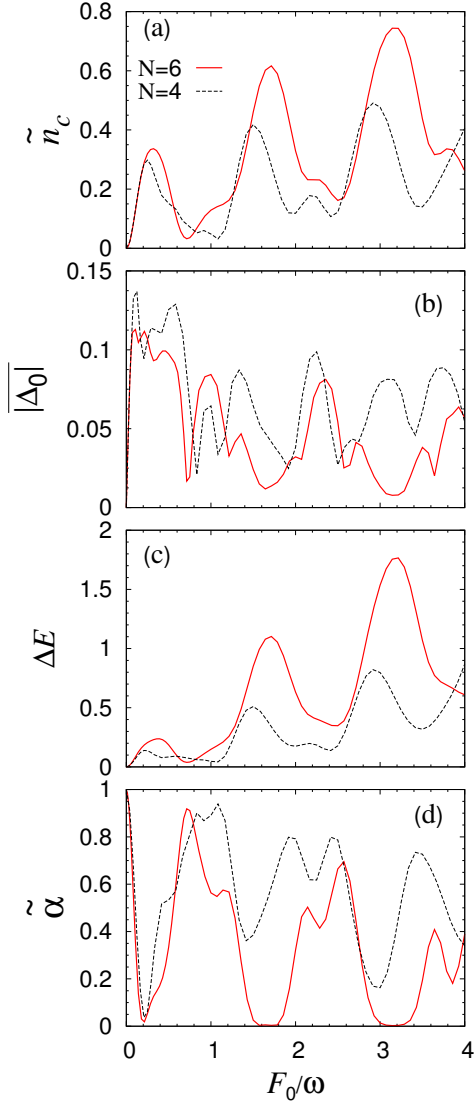}
\caption{Same plots as Fig. \ref{fig:lanczos-time-av-gauss} by the ED method except that we use 
$F(\tau)$ with the rectangular envelope.}
\label{fig:lanczos-time-av-sin}
\end{figure}

In Fig. \ref{fig:lanczos-time-av-sin}, we show $\widetilde{n_c}$, $\overline{|\Delta_0|}$, 
$\Delta E$, and $\widetilde{\alpha}$ as functions of $F_0/\omega$ obtained by the 
ED method when we use $F(\tau)$ with the rectangular envelope. 
The parameters are the same as those in Fig. \ref{fig:lanczos-time-av-gauss}. 
For $F_0/\omega\lesssim 1$, the $F_0/\omega$ dependence of these quantities is similar 
to those in Fig. \ref{fig:lanczos-time-av-gauss}, indicating that the pulse shape 
does not significantly affect our results as in the case of the HF method. 
The cyclic behavior is 
evident even for $F_0/\omega>1$, although in this region the increase (decrease) in 
$\widetilde{n_c}$ and $\Delta E$ ($\widetilde{\alpha}$) does not correspond to the 
large enhancement in $\overline{|\Delta_0|}$, which is in contrast to the results with 
the HF method shown in Fig. \ref{fig:fig11}. This is caused by the dephasing
 discussed in Sect. V, which suppresses $|\Delta_0|$. In fact, 
for $F_0/\omega\sim 1.6$ and 
$3.2$ of the results with $N=6$, where $\widetilde{n_c}$ and $\Delta E$ exhibit 
a peak and $\widetilde{\alpha}\sim 0$, 
we have confirmed that the time profile of $|\Delta_0|$ is similar to that in Fig. 
\ref{fig:lanczos-time-profile-gauss-large}.

\begin{acknowledgments}
This work was supported by JSPS KAKENHI Grant Nos. JP15H02100, JP16K05459, 
JP19K23427, and JP20K03841, MEXT Q-LEAP Grant No. JPMXS0118067426, 
JST CREST Grant No. JPMJCR1901, and Waseda University Grant for Special 
Research Projects (Project No. 2020C-280). 
\end{acknowledgments}


\begin{thebibliography}{99}
\bibitem{comment1}
The assumption that $\langle c^{\dagger}_{i\sigma}f_{i\sigma}\rangle$ and 
$\langle n^c_{i\sigma}\rangle$ are independent of $i$ is in fact validated by 
performing unrestricted HF calculations in 
real space that can treat inhomogeneous mean-field solutions.
\bibitem{comment2}
If the initial order parameters are independent of $i$, the expectation values like 
$\langle c^{\dagger}_{k\sigma}f_{k+{\bm Q}\sigma}\rangle$ 
and $\langle c^{\dagger}_{k\sigma}c_{k+{\bm Q}\sigma}\rangle$ with ${\bm Q}\neq 0$ are always
 zero during the mean-field dynamics. This can be verified by using the equations of motion for these 
expectation values under the Hamiltonian Eq. (\ref{eq:ham_mat}). Therefore, 
$\langle c^{\dagger}_{i\sigma}f_{i\sigma}\rangle$ and 
$\langle n^c_{i\sigma}\rangle$ are independent of $i$ at any $\tau>0$.
\end{thebibliography}
\end{document}